\documentclass[sigconf]{acmart}
\usepackage{acronym}
\acrodef{AI}{Artificial Intelligence}
\acrodef{LLMs}{Large Language Models}
\acrodef{ACTA}{Applied Cognitive Task Analysis}

\AtBeginDocument{%
  \providecommand\BibTeX{{%
    \normalfont B\kern-0.5em{\scshape i\kern-0.25em b}\kern-0.8em\TeX}}}

\usepackage{multirow}
\usepackage{enumitem}
\usepackage{subcaption}
\usepackage{listings}
\usepackage{xcolor}

\begin{document}

\title[From Junior to Senior]{From Junior to Senior: Allocating Agency and Navigating Professional Growth in Agentic AI–Mediated Software Engineering}

\author{Dana Feng}
\authornote{Feng and Yun contributed equally to this work  (order was decided by last coin toss)}
\affiliation{%
  \institution{Independent Researcher}
  \city{Brooklyn}
  \state{NY}
  \country{USA}
}
\email{danafeng308@gmail.com}

\author{Bhada Yun}
\authornotemark[1]
\affiliation{%
  \institution{ETH Z{\"u}rich}
  \city{Z{\"u}rich}
  \country{Switzerland}
}
\email{bhayun@ethz.ch}

\author{April Yi Wang}
\affiliation{%
  \institution{ETH Z{\"u}rich}
  \city{Z{\"u}rich}
  \country{Switzerland}
}
\email{april.wang@inf.ethz.ch}

\renewcommand{\shortauthors}{Feng and Yun et al.}



\begin{abstract}
Juniors enter as AI‑natives, seniors adapted mid‑career. AI is not just changing how engineers code---it is reshaping who holds agency across work and professional growth. We contribute junior–senior accounts on their usage of agentic AI through a three-phase mixed-methods study: ACTA combined with a Delphi process with 5 seniors, an AI-assisted debugging task with 10 juniors, and blind reviews of junior prompt histories by 5 more seniors. We found that agency in software engineering is primarily constrained by organizational policies rather than individual preferences, with experienced developers maintaining control through detailed delegation while novices struggle between over-reliance and cautious avoidance. Seniors leverage pre-AI foundational instincts to steer modern tools and possess valuable perspectives for mentoring juniors in their early AI-encouraged career development. From synthesis of results, we suggest three practices that focus on preserving agency in software engineering for coding, learning, and mentorship, especially as AI grows increasingly autonomous.
\end{abstract}

\begin{teaserfigure}
    \centering
    \includegraphics[width=\linewidth]{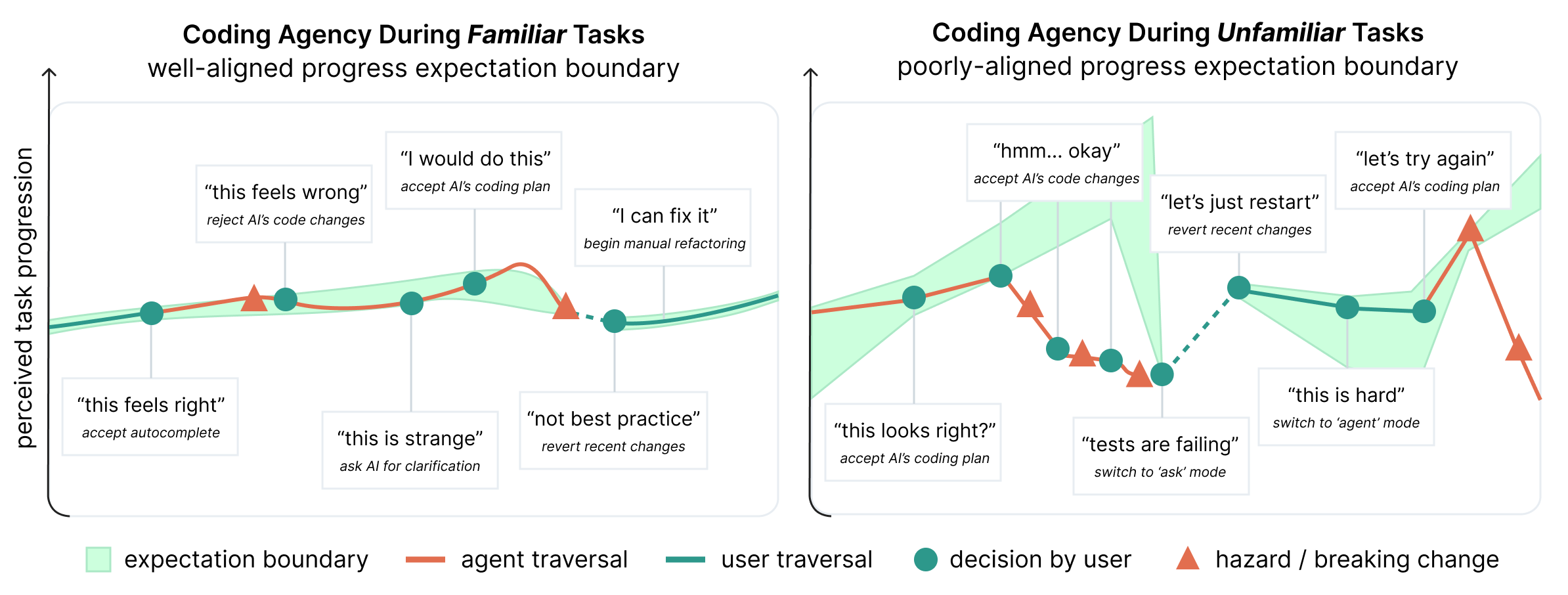}
    \caption{\textbf{Agency allocation patterns in AI-mediated software engineering across task familiarity.} \textbf{Left} panel shows coding during familiar tasks with well-defined expectation boundaries, where both groups maintain control through strategic accept/reject decisions. \textbf{Right} panel depicts unfamiliar tasks with poorly-defined, shifting expectation boundaries, revealing divergent agency patterns when using agentic AI tools. Green paths indicate user-maintained control, orange paths show agent-driven traversal, with circles marking key decision points and triangles representing potential hazards or breaking changes.}
    \Description{Two-panel diagram comparing agency allocation in familiar versus unfamiliar coding tasks. Each panel shows task progression over time with expectation boundaries, decision points, and traversal paths. Left panel demonstrates stable boundaries and controlled interactions. Right panel shows shifting boundaries and more complex navigation patterns, including mode switching between agent and ask modes.}
    \label{fig:teaser}
\end{teaserfigure}




\settopmatter{printacmref=false} 
\setcopyright{none}
\renewcommand\footnotetextcopyrightpermission[1]{} 
\pagestyle{plain} 

\maketitle

\section{Introduction}

\begin{quote}
``\dots the notion of the programmer as an easily replaceable component in the program production activity has to be abandoned. Instead the programmer must be regarded as a responsible developer and manager of the activity in which the computer is a part.'' \cite{naur1985programming} \\- Peter Naur, 1985
\end{quote}

\emph{Implement the feature, add tests, open the PR} used to be an engineer's checklist. Today, \ac{AI} agents can carry out all three, sometimes better, sometimes worse, almost always faster, until it's not \cite{recupito2024technical}. Industry momentum has been brisk: ``vibe coding'' has gone mainstream; repositories ship agent metadata (e.g., \texttt{AGENTS.md}) to steer AI toolchains \cite{agentsmd2025}; model providers advertise ``learning modes'' that retain context across sessions \cite{openai2025study}; and some firms are experimenting with AI-enabled interviews that explicitly permit tool use \cite{koebler2025meta}. Concerns have shifted over time: from human error in code, to AI hallucinations, and now to delegation, responsibility, and control when capable agents are part of the workflow. The distinction between generative AI (which produces new content through prompt-response interaction) and agentic AI (which autonomously executes actions) becomes crucial as developers increasingly adopt tools like Cursor and GitHub Copilot that can independently modify and interpret codebases \cite{li2025rise}.

Recent HCI and SE research suggests that AI will become increasingly integrated into software development, though the partnership remains symbiotic: AI serves as a tool, while human judgment, ingenuity, and creativity are essential to evaluate and complement AI output \cite{10.1145/3715003, abrahao2025software, 10.1145/3709360}. Thus, it is critical to understand how agency allocation---the distribution of decision authority and accountability---with AI is distributed across experience levels. In industry, junior engineers enter the workforce alongside AI, navigating the tension between productivity and learning, while senior engineers must adapt mid-career, revising their leadership and mentorship practices. Investigating this generational divide is particularly important because traditional mechanisms for transferring and building expertise---such as pair programming, discussion, and documentation \cite{mtsweni2018issues, ryan2013acquiring}---now need to integrate AI as both a tool and an intermediary.

Software engineering has always depended on a talent pipeline where juniors gradually mature into seniors by developing best practices, judgment under uncertainty, and mentoring capacities. AI unsettles this process, especially as industry narratives diverge: some firms are cutting junior roles as ``replaceable by AI,'' while others prize ``AI-native'' hires \cite{npr2025jobs, wsj2025aisalaries, jockimsAI, scheiberWhich}. Prior work has highlighted the importance of investigating both the technical and soft skills required for AI-augmented software engineers \cite{abrahao2025software}, as well as the effects on team dynamics \cite{Pezze2025Roadmap2030}. Building on this, it is critical to understand how professional growth, learning, and mentoring align with idealized human-AI collaboration---particularly when companies take extreme approaches to AI adoption.

In this study, we zoom in on AI‑native junior engineers and senior engineers with pre‑AI instincts. We conducted a three-phase qualitative study with junior and senior software engineers to investigate how generative and agentic AI tools shape software practice, professional growth, and mentorship. Our approach combined semi-structured interviews, structured elicitation techniques (\ac{ACTA} \cite{militello1998applied}, Delphi method \cite{ hasson2000research, goodman1987delphi}), and task-based activities (AI-assisted debugging and prompt review). 
In Phase 1, we interviewed five senior engineers with ACTA to elicit examples of tacit knowledge \cite{son2024demystifying, chin2021easier}, refining these through Delphi-inspired consensus-building to converge on a final, realistic debugging task. In Phase 2, 10 junior engineers engaged in the Phase 1 final task using Cursor, an agentic AI tool, followed by postmortems, surveys, and reflective interviews on their AI use. In Phase 3, five different senior engineers reviewed anonymized junior artifacts---including code, prompt histories, and reflections---mirroring realistic mentorship contexts and evaluating how AI records support knowledge transfer.

Across these phases, we also conducted semi-structured interviews, enabling us to triangulate how engineers allocate agency with AI, perceive career development, and approach mentorship or learning in AI-mediated workflows.

Our findings highlight that agency allocation in AI-mediated software work is preconfigured at the organizational layer (policies, tooling defaults, repos, CI guardrails) before individual preferences matter. Within those bounds, seniors and juniors take varying routes to preserve control with the way they engage with generative and agentic AI. 
Both groups use small, task-focused prompts in high familiarity contexts, while in low-familiarity contexts, juniors often over-rely on agents or correct hallucinations with limited intuition, while seniors use AI more strategically for idea generation or low-level tasks. Both emphasized responsibility for their code, and also voiced concerns for skill degradation and lack of code understanding. Seniors framed growth around trial-and-error, strong mental models, and soft skills, and juniors reported impostor syndrome and feelings of limited accomplishment. In addition to code output and judgment, agency in the form of learning and mentorship also manifested: seniors positioned themselves as leaders who contextualize and critique AI outputs, and juniors valued AI for basic questions but still relied on human discussion to build deeper contextual understanding.

Our work highlights opportunities for three evolving practices in the AI-mediated software engineering industry:

\begin{itemize}[topsep=0pt, partopsep=0pt, parsep=0pt, itemsep=0.5em]
    \item \textbf{Preserving Individual Agency:} Company-wide and personal practices for using AI tools---such as incremental changes and interrupting and verifying outputs---help engineers maintain control and responsibility over their work.
    
    \item \textbf{Evolving the Mentorship Pipeline:} Collaborative and individual practices where senior engineers actively transmit intuition, critical thinking, and judgment to help junior engineers retain agency over their learning and professional growth in an AI-mediated environment.
    
    \item \textbf{Prompt \& Code Reviews (PCRs):} Collaborative practices that structure AI interactions to uphold accountability---juniors document and justify key prompts, while seniors oversee the process and ensure juniors remain authors of their reasoning and retain ownership of outputs.
\end{itemize}

\section{Related Work}

\subsection{Generative/Agentic AI in the Software Industry}
\ac{LLMs}, a type of Generative AI, produce new content typically accessed through a prompt-and-response interaction. Agentic AI, however, goes a step further: rather than leaving the programmer to act on an output, it can autonomously execute actions in response to a prompt, with the programmer primarily reviewing, approving, or rejecting changes. Put differently, agents are autonomous, iterative systems that perceive feedback and operate in dynamic environments, whereas a standalone LLM is not, by itself, an agent. Consequently, software engineers are increasingly integrating agentic AI into their workflows when compatible \cite{chen2025need, Lambiase2024BotsMLR}. Popular examples---such as GitHub Copilot, Cursor AI, and Windsurf \cite{github_copilot, cursor_agents, windsurf_ide}---seamlessly embed into developer IDEs \cite{li2025rise}.
Agentic AI also supports ``vibe coding,'' where high-level natural language prompts lead to the evaluation, generation, and deployment of code \cite{Ray2025VibeCodingReview}, shifting greater control to the agentic system and reducing the user's control \cite{Sapkota2025VibeAgentic}.

Prior research has examined how both software engineers and students engage with generative and agentic AI across a range of contexts. Studies of expert developers have explored how they use AI to perform software tasks \cite{yen2024coladder, kumar2025sharp, zamfirescu2025beyond}, while investigations of junior engineers highlight challenges such as reduced learning opportunities alongside productivity gains \cite{schmitt2024generative}. Studies have also surfaced students' and professionals' interactions with coding agents across experience levels \cite{chen2025need, barke2023grounded}, and suggested design guidelines and collaboration frameworks to enhance AI use \cite{10.1145/3696630.3728714, 10.1145/3643690.3648236}. Building on this foundation, recent work recommends further exploratory studies to understand how groups with different levels of expertise experience and take part in AI adoption \cite{ferino2025junior}, especially as agentic AI tools evolve and become increasingly widespread.

Thus, we extend prior work by investigating how junior and senior engineers engage with both generative and agentic AI in their day to day jobs. Our study looks at two ends of the spectrum: engineers who began their careers alongside AI and those who spent most of their careers without it. Rather than detailing specific workflows, we focus on how these groups preserve control, balance productivity and learning, and experience psychosocial impacts such as imposter syndrome.

\subsection{Agency for Software Engineers}
Broadly, agency is defined as the ability to act driven primarily by internal thoughts and feelings, rather than the external environment \cite{dennett1978brainstorms, Trafton2024PerceptionOfAgency}. In a professional sense, Eteläpelto et al. defines professional agency as exercised when ``professional subjects and/or communities influence, make choices, and take stances on their work and professional identities'' \cite{Etelapelto2013ProfessionalAgency}. In the context of software engineering with AI, we are interested in the agency of the AI tool as well as the human engineer's stance on their own professional identity when using the AI tool, and who is to blame for the decisions made by AI tools \cite{Heaton2023SocialImpactADM}. Heer argues that AI systems should augment rather than replace people, safeguarding human agency, and that the user should remain the ultimate decision-maker \cite{Heer2019AgencyPlusAutomation}. The use of AI brings up the issue of negotiating human control for agentic autonomy; some advocate for a hybrid future where natural language ideation guides autonomous execution \cite{Sapkota2025VibeAgentic}, while others question if we have to use AI on every task \cite{Xia2025Agentless}. Prior work has examined agency across various domains. In social media, researchers explored the tension between algorithmic curation of content and users' control over video editing features \cite{KangLou2022TikTokAgency}. In design contexts, studies highlighted how AI is perceived either as a tool or as a collaborator within the creative process \cite{Lawton2023WhenIsATool}. Building on these insights, we investigate software engineering to understand when and how engineers exercise agency in using AI, and under what circumstances that agency may be diminished. As found by Nam, it is imperative to research developers' motivations and reasons for code completions with LLM for future software tool designs, especially since professionals tend to directly use LLM prompts \cite{nam2024using}. Thus, our study not only focuses on practical and realistic tasks with agentic AI (Cursor AI debugging and prompt reviews), but also explores the motivators and inhibitors when it comes to using AI for both junior and senior engineers across different companies in industry.

\subsection{Professional Growth, Learning, and Mentorship in Software Engineering}
Tacit knowledge is knowledge that is difficult to articulate or transfer \cite{son2024demystifying}, ranging from understanding company-specific software and the rationale behind program design \cite{naur1985programming} to intuition about where to focus when debugging. Pezzè et al. note that, unlike human-generated code---where accountability can be implicitly assigned---automatically generated code introduces issues of trust, requiring high confidence and evidence of correctness \cite{Pezze2025Roadmap2030}, much of which relies on tacit knowledge. Ensuring, validating, and refining generated code thus demands extensive tacit expertise, particularly when resolving ambiguities \cite{Lyu2025AutomaticProgramming}. Empirical studies have examined aspects of tacit knowledge in software engineering, such as ``API documentation smells,'' which hinder clarity despite being technically correct \cite{Khan2021AutomaticDetectionOfFiveAPIDocumentationSmells}, and there have been efforts to externalize tacit knowledge in software systems \cite{johanssen2019tacit, bjornson2008knowledge, peng2025code}. Prior work also highlights AI's potential to teach essential soft skills for software engineering \cite{fox2025using} and explores mentorship in both academic and professional contexts \cite{hang2024industry, ko2017computing, levine2022improving}. Building on this foundation, our study examines how senior engineers' tacit knowledge informs career growth, how mentorship manifests in industry, and where AI can do the teaching versus when guidance from a senior remains indispensable.

With the growing use of AI in industry, HCI and SE research has emphasized preparing software students for AI-supported workflows \cite{10.1145/3639474.3640061, 10.1145/3626252.3630927}. Consequently, studies have explored leveraging AI to teach the software development life cycle (SDLC) \cite{wang2025devcoach} and benchmarking multiple generative AI tools across common software engineering learning tasks \cite{roy2025benchmarking}. Learning, however, extends beyond school; in industry, engineers must balance acquiring knowledge with maintaining productivity. One key avenue for learning is code review, which serves not only to ensure code quality but also to facilitate knowledge sharing across team and seniority levels \cite{10.1145/3474624.3477063}. Gao et al. highlight how LLMs are transforming code review by automating the review process, though challenges of applying industry specific practices and evaluating purposes continue to persist \cite{gao2024currentllm4se}. Systems such as Meta-Manager have also been developed to track code provenance to answer developer questions about an unfamiliar codebase \cite{horvath2024meta}. Building on this, we examine not only how junior and senior engineers learn with AI, but also the role of human prompt review in supporting both quality assurance and mentorship in code creation and generation.
\section{Method Overview}

We designed a three-phase study with both senior and junior software engineers, combining semi-structured interviews, structured elicitation techniques (Delphi Method, ACTA), and task based activities (Cursor debugging task and prompt reviews).
With these methods, we aimed to answer the following research questions:

\begin{enumerate}
    \item \textbf{RQ1}: How do junior and senior software engineers allocate agency \footnote{By `agency allocation' we mean who holds decision authority and who is answerable for reasons at each task stage.} between themselves and agentic / generative AI in daily work?
    \item \textbf{RQ2}: How do junior and senior software engineers perceive professional growth for a junior in the age of AI?
    \item \textbf{RQ3}: When and why do engineers deem mentorship indispensable in workflows with agentic / generative AI?
    \item \textbf{RQ4}: How do AI records (e.g., prompt history, provenance) shape code review and mentorship?
\end{enumerate}

\subsection{Participants}

We recruited \textbf{20} professional software engineers (10 juniors, 10 seniors) across three phases using convenience and snowball sampling (LinkedIn outreach and participant referrals). We defined \emph{seniors} as engineers with $\geq$5 years full-time experience and $\geq$1 year in an advanced role (senior, staff, or lead), and \emph{juniors} as engineers with $\leq$1 year full-time experience. Recruitment targeted a spread of company types (e.g., Big Tech/Cloud, FinTech/Financial Services, Enterprise SaaS, DevTools, HealthTech) to capture variability in AI policies and tooling. Our goal was \emph{qualitative, exploratory insight} across 1-hour long semi-structured and task based interviews into how agency and mentorship shift with agentic AI across experience levels. We therefore prioritized \emph{depth over breadth} in a three-phase design detailed below.

\begin{table*}[htbp]
    \centering
    \small
    \caption{\textbf{Phase 1 (S1--S5) participants: experience, roles, and AI tool usage.}}
    \Description{Tabular summary of five senior software engineers S1 to S5 with columns: ID, SWE experience and senior experience, company type, title, generative AI tools with frequency and duration, agentic AI tools with frequency and duration, and self-reported familiarity with agentic AI on a 1 to 5 scale.}
    \label{tab:P1}
    \renewcommand{\arraystretch}{1.5}
    \begin{tabular*}{\textwidth}{l p{0.6in} p{0.9in} p{1.2in} p{1in} p{1.24in} p{1in}}
        \toprule
        ID & Total/Sr. Exp. & Company Type & Title & GenAI usage, frequency, and duration & Agentic AI usage, frequency, and duration & Agentic AI familiarity \\
        \midrule
        S1 & 9 yrs / 1.5 & Big Tech / Cloud \& AI & Sr Software Engineer & GPT, Gemini daily 1.5 yr & Cline, Cursor few/wk 6 mo & somewhat familiar \\
        S2 & 12 yrs / 6 & FinTech & Lead Software Engineer & Gemini few/mo 1 yr & Cursor, Augment (not at work) & familiar \\
        S3 & 5 yrs / 3 & Design Tools & Senior Applied Scientist & Claude, GPT daily 1 yr & Cursor, Claude Code few/wk 1 yr & familiar \\
        S4 & 13 yrs / 2 & DevTools startup & Senior Software Engineer & Claude Desktop daily 3 mo & Cursor, Claude Desktop daily 3 mo & somewhat familiar \\
        S5 & 13 yrs / 4 & Enterprise SaaS & Lead Developer & Gemini daily 2--3 mo & Cursor <1/mo 2--3 mo & not very familiar \\
        \bottomrule
    \end{tabular*}
\end{table*}

\begin{table*}[h]
    \centering
    \small
    \caption{\textbf{Phase 2 (J1--J10) participants: experience, roles, and AI tool usage.}}
    \Description{Tabular summary of ten junior software engineers J1 to J10 with columns: ID, SWE experience, company type, title, generative AI tools with frequency and duration, agentic AI tools with frequency and duration, and self-reported familiarity with agentic AI on a 1 to 5 scale.}
    \label{tab:P2}
    \renewcommand{\arraystretch}{1.5}
    \begin{tabular*}{\textwidth}{l p{0.6in} p{0.9in} p{1in} p{1.17in} p{1.24in} p{1in}}
        \toprule
        ID & SWE Exp. & Company Type & Title & GenAI usage, frequency, and duration & Agentic AI usage, frequency, and duration & Agentic AI familiarity \\
        \midrule
        J1 & 1 yr & FinServ startup & Software Engineer & GPT, Claude daily 1 yr & Gemini reviewer, Cursor multi/day 10 mo & familiar \\
        J2 & 1 yr & FinTech & Software Engineer & Copilot daily 1 yr & Copilot multi/day 1 yr & somewhat familiar \\
        J3 & 1 yr & Enterprise SaaS & Software Engineer & GPT, Claude few/mo 2 yr & Cursor multi/day 9 mo & very familiar \\
        J4 & 1 yr & HealthTech & Software Engineer & GPT few/wk 1 yr & Copilot, Cursor multi/day 1 yr & familiar \\
        J5 & <1 yr & Big Tech / Cloud \& AI & Software Engineer & GPT, Gemini daily 1 yr & company version of Cursor few/wk 2 mo & somewhat familiar \\
        J6 & 1 yr & FinTech & Software Engineer & GPT, Claude few/mo 1 yr & Cursor few/mo 6 mo & not very familiar \\
        J7 & 1 yr & FinMedia & Software Engineer & GPT, Claude <1/mo 2 mo & Copilot <1/mo 2 mo & not very familiar \\
        J8 & 1 yr & Investment & Software Engineer & GPT daily 1 yr & Copilot daily 1 yr & never used \\
        J9 & 9 mo & Payments & Software Engineer & GPT few/wk 9 mo & Copilot, Cursor daily 9 mo & somewhat familiar \\
        J10 & 1 yr & Consulting & Software Engineer I & GPT, Canva AI daily 1 yr & Copilot few/wk 1 yr & familiar \\
        \bottomrule
    \end{tabular*}
\end{table*}

\begin{table*}
    \centering
    \small
    \caption{\textbf{Phase 3 (S6--S10) participants: experience, roles, and AI tool usage.}}
\Description{Tabular summary of senior software engineers S6 to S10 with columns: ID, SWE experience and senior experience, company type, title, generative AI tools with frequency and duration, agentic AI tools with frequency and duration, and self-reported familiarity with agentic AI on a 1 to 5 scale.}
\label{tab:P1b}
\renewcommand{\arraystretch}{1.5}
    \begin{tabular*}{\textwidth}{l p{0.6in} p{0.9in} p{1.2in} p{1in} p{1.19in} p{1in}}
    \toprule
    ID & Total/Sr. Exp. & Company Type & Title & GenAI usage, frequency, and duration & Agentic AI usage, frequency, and duration & Agentic AI familiarity \\
    \midrule
        S6 & 25 yrs / 12 & FinTech & Lead Software Engineer & GPT, Claude few/wk 2 yr & Copilot few/wk 2 yr & not very familiar \\
        S7 & 13 yrs / 8 & Social / Consumer & Staff Software Engineer & GPT few/mo 6 mo & Copilot few/mo 6 mo & somewhat familiar \\
        S8 & 18 yrs / 11 & Big Tech / Cloud \& AI & Staff Software Engineer & GPT, Claude daily 2 yr & Gemini few/wk 1.5 yr & somewhat familiar \\
        S9 & 26 yrs / 16 & Insurance / Health & Senior Software Developer & Copilot Chat daily 1 yr & Copilot, Gitlab Duo few/wk 1 yr & not very familiar \\
        S10 & 8 yrs / 6 & FinTech & Senior Software Engineer & Gemini few/wk 1.5 yr & Copilot <1/mo 1.5 yr & somewhat familiar \\
        \bottomrule
    \end{tabular*}
\end{table*}

\subsection{All Phases}

\subsubsection{Pre-Interview Survey}
Prior to interviews, all participants completed a survey capturing demographic information,
years of professional experience, and details regarding the types and usage of AI-assisted tools.

\subsection{Phase 1 (P1)}

We conducted 60-minute interviews with 5 senior software engineers (all male, aged 28–46, based in the US), supplemented by one survey, all administered remotely via Zoom.

\subsubsection{Semi-Structured Interview - 20 minutes}
We introduced a working definition of tacit knowledge and asked engineers to reflect on how they had acquired tacit expertise, how AI-assisted coding tools and IDEs (e.g., GitHub Copilot, ChatGPT, Cursor) affect their workflows, and their mentorship practices with AI.

\subsubsection{Task-Based Elicitation - 40 minutes}
To capture concrete examples of tacit expertise, we integrated task-based elicitation into the interviews. Using the ACTA framework (Task diagram, Knowledge Audit, Simulation), participants decomposed domain-relevant scenarios (e.g., debugging or feature development) into steps. For each step, they identified cognitively demanding judgments and reflected on the risks and benefits of delegating these steps to AI, some of which are shown in Appendix~\ref{sec:acta-brainstorm}. This approach surfaced ``hidden'' reasoning strategies often overlooked in interviews alone.

We employed ACTA \cite{militello1998applied} to systematically elicit the implicit (i.e. tacit) knowledge and cognitive strategies that distinguish senior from junior engineers---expertise developed over years that shapes how seniors approach complex problems differently. While traditional structured interviews might capture what engineers do, ACTA's structured task decomposition reveals how they think through problems, identify critical decision points, and recognize patterns that novices often miss. This method was particularly valuable for generating authentic debugging scenarios for Phase 2, as it surfaces the ``hidden'' cognitive demands within routine tasks that seniors navigate intuitively but juniors may struggle with, especially when relying on AI assistance. By having seniors explicitly articulate which aspects of tasks require judgment versus which could be delegated to AI, ACTA helped us identify tasks that would reveal meaningful differences in how juniors and seniors allocate agency (RQ1) and approach problem-solving with AI tools. 

\subsubsection{Consensus-Building (Delphi-Inspired) Post-Interview Survey}
We then used a Delphi-inspired process to rank the elicited examples. Task descriptions from earlier sessions were anonymized (e.g., company-specific tools and libraries were replaced) and edited into more structured, task-like formats by the full-time industry software engineer on the author team (e.g., by clarifying task components and adding solutions mentioned during the interviews). Task descriptions were also consolidated across sessions when they were very short or similar, including a case where different parts of the single task was drawn from multiple sessions. The tasks were then circulated among participants via a survey for evaluation, where seniors reviewed and rated each task, suggesting edits, comments, or context to improve the task. The final task was chosen according to participant rankings, with four engineers rating it as a 4 and one engineer rating it as a 5 in terms of its importance for demonstrating tacit knowledge.

We complemented ACTA with a Delphi-inspired consensus-building process \cite{hasson2000research, goodman1987delphi} to ensure the final debugging task reflected collective expert judgment rather than researcher assumptions or individual biases. Rather than arbitrarily selecting a task ourselves, we leveraged the distributed expertise of practicing senior engineers to evaluate, refine, and rank the elicited scenarios based on their real-world importance for demonstrating the experience gap between junior and senior engineers. 

\subsection{Phase 2 (P2)}
We conducted 65-minute interviews with 10 junior engineers (4 female, 6 male, aged 22–24, based in the US), supplemented by one survey, all administered remotely via Zoom.

\subsubsection{Task Setup (5 minutes)}
\begin{figure}[h]
    \begin{subfigure}[b]{0.48\textwidth}
        \centering
        \includegraphics[width=\textwidth]{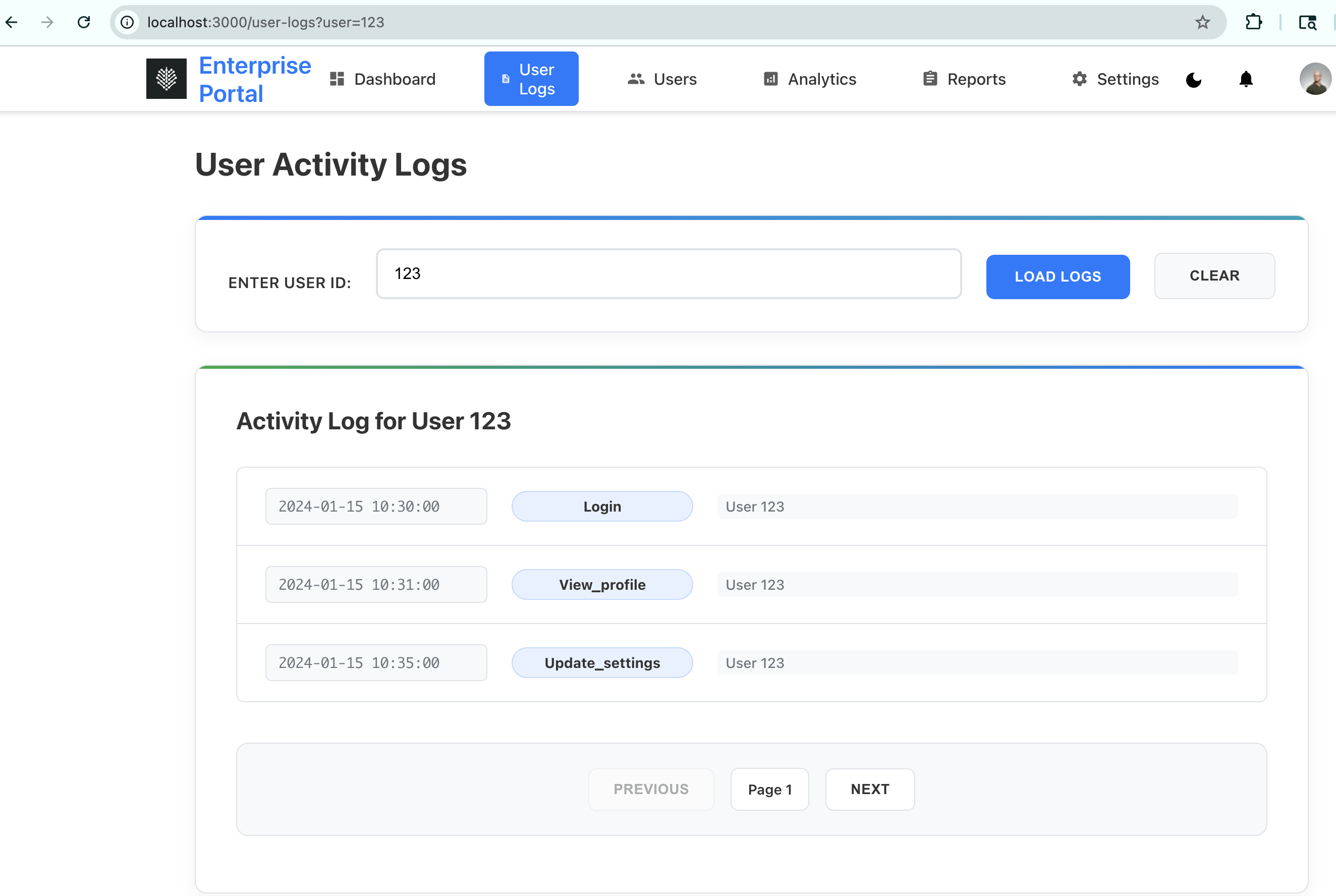}
        \caption{User Activity Logs page}
        \label{fig:admin2}
    \end{subfigure}
    \hfill
    \begin{subfigure}[b]{0.48\textwidth}
        \centering
        \includegraphics[width=\textwidth]{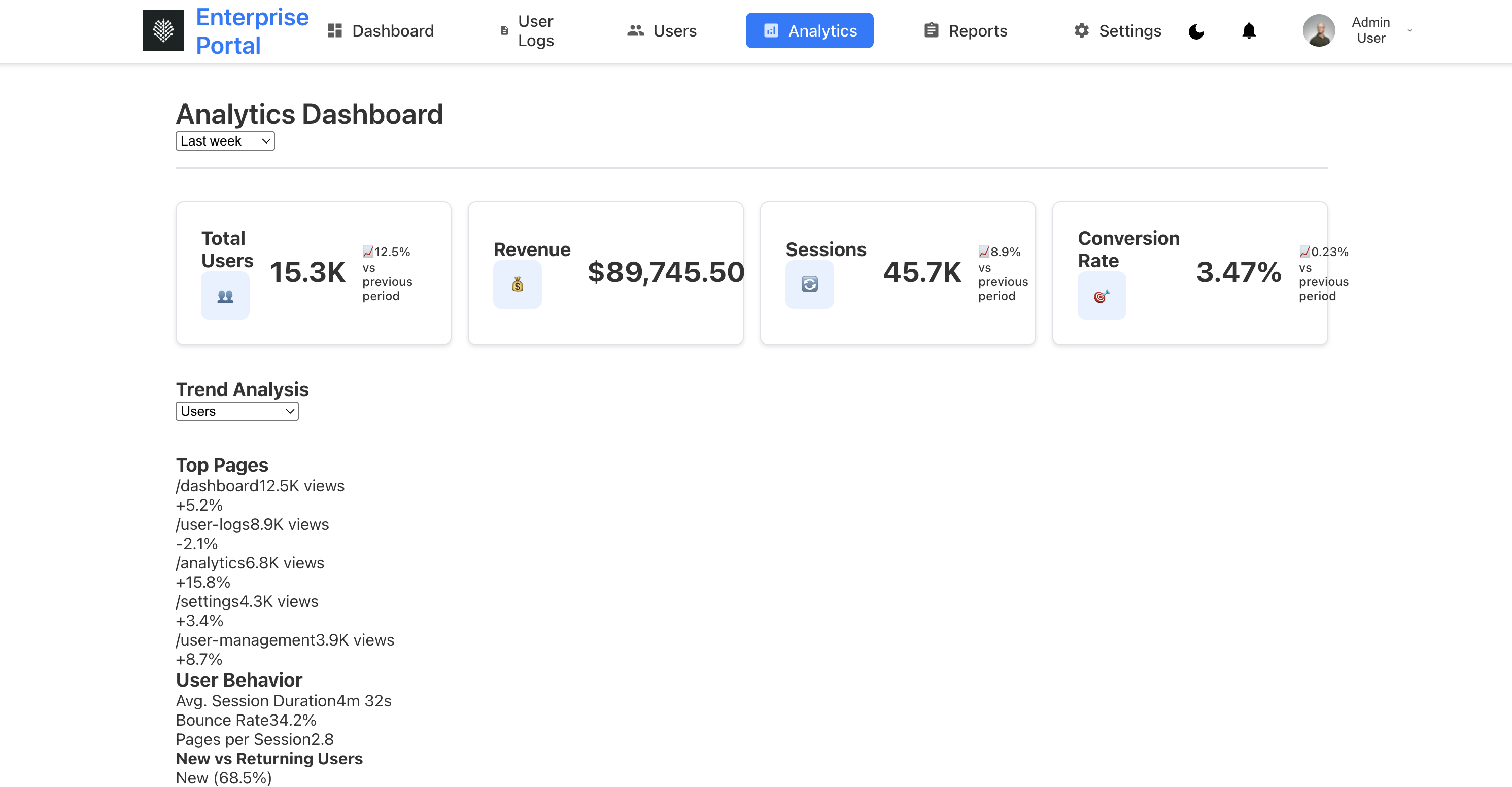}
        \caption{Analytics Dashboard}
        \label{fig:admin3}
    \end{subfigure}
    \caption{\textbf{Debugging task interface showing the React-based admin panel application with multiple bugs across different pages.} The User Activity Logs page (a) displays user authentication and activity tracking with pagination controls. The Analytics Dashboard (b) presents key metrics (users, revenue, sessions, conversion rate), trend analysis, top page performance, and user behavior statistics.}
    \Description{Two screenshots of the React-based admin panel application. The top screenshot shows the User Activity Logs page with user ID input, load logs button, and activity entries including login, profile view, and update settings actions with timestamps. The bottom screenshot shows the Analytics Dashboard with four metric cards displaying total users (15.3K), revenue (\$89,745.50), sessions (45.7K), and conversion rate (3.47\%), followed by a trend analysis dropdown, top pages list with view counts and percentage changes, and user behavior metrics.}
    \label{fig:admin}
\end{figure}

Based on the P1 final tacit knowledge task, we developed a React-based administrative application featuring multiple pages, including a User Activity Logs page and an Analytics Dashboard (Figure \ref{fig:admin}). Participants installed Cursor, cloned the task repository, installed dependencies, and launched the application. 
To simulate scenarios where errors may not be immediately visible to developers, console errors were suppressed and participants used incognito mode for standardization. This setup reflects realistic challenges, such as debugging an infinite loop in a React app that occurs silently in production due to React suppressing errors, as recounted by a participant in P1.

\subsubsection{Main Task (30 minutes)}
Participants engaged in a realistic maintenance scenario detailed below. They were permitted to use Cursor in either \emph{Agent} mode (autonomously completing coding tasks) or \emph{Ask} mode (scanning the codebase and answering questions) according to their normal workflow. It is important to note that both modes are considered agentic, as they both rely on the agent autonomously searching the codebase and making decisions. No think-aloud protocol was required to ensure a more natural workflow \cite{richardson2017think}; instead, we collected participants' code, prompts, and post-hoc reflections.  

\emph{Task Details:} The task involved debugging a React-based admin panel application that hangs in production without showing errors. The application contained three distinct bugs: (1) incorrect service worker routing for query-parameterized URLs, (2) nested \texttt{useEffect} hooks causing infinite re-renders, and (3) improper use of \texttt{window.href.location} resulting in duplicate loading pages. This three part task was chosen because senior engineers noted that it requires juniors to apply tacit knowledge of React effects and service workers in a realistic debugging scenario. The issue had no console errors, so identifying the problem relies on intuition, trial-and-error, and careful observation of software output---skills that AI cannot easily replicate. The task balances a clear initial step with enough complexity to reveal a junior's problem-solving, understanding of React, and ability to reason through production-like challenges.

\subsubsection{Documentation (10 minutes)}
Participants produced a postmortem summarizing the bug, the diagnostic process, and recommended prevention strategies. Participants were verbally instructed not to use AI tools during this phase; this guidance was not part of the written protocol and was introduced after the first interview. During that initial session, the participant used Cursor to generate a draft postmortem, which they subsequently edited and refined. To ensure that later participants reflected on---and fully understood---their written responses, the procedure was adjusted to discourage AI assistance. All subsequent participants followed the guidance and did not use Cursor for the postmortem. The first participant's use of Cursor did not affect the analysis reported in this paper, as we do not focus on postmortem writing.

\subsubsection{Semi-Structured Interviews (20 minutes)}
Semi-structured interviews explored participants' experiences with the task, their interactions with AI, patterns of overreliance and learning, and the professional, collaborative, and emotional implications of AI usage.

\subsubsection{Post-Task Survey}
Participants completed the NASA-TLX and SMEQ workload scales \cite{hart1988development, sauro2009comparison}, in addition to retrospective questions evaluating confidence and perceived learning.

\subsection{Phase 3 (P3)}

In the final phase, we conducted 60-minute interviews with 5 senior software engineers (all male, aged 31-55, based in the US or Canada), all administered remotely via Zoom. Each senior was randomly assigned two unique juniors, with no junior reviewed more than once.

\subsubsection{Semi-Structured Interviews (20 minutes)}
Senior engineers discussed how AI-assisted coding tools (e.g., GitHub Copilot, ChatGPT, Cursor) influence their workflows, as well as their mentorship practices involving AI.

\subsubsection{Codebase Familiarization and Task Introduction (6 minutes)}
Participants were guided through the three bugs present in the codebase and provided with a brief overview of the task assigned to the junior engineers.

\subsubsection{Evaluation of First Junior's Artifacts (12 minutes)}
Seniors reviewed the anonymized artifacts of one junior participant, including:  
\begin{itemize}
    \item \emph{Code and Postmortem} (pull request and reflective summary).  
    \item \emph{Prompt History} (prompts, answers, and mode switches).  
    \item \emph{Interaction Statistics} (task duration, acceptance or rejection of AI suggestions).  
\end{itemize}
They assessed code quality, reasoning processes, and identified both accurate and erroneous assumptions via live cognitive walkthroughs. This setup reflects realistic mentorship constraints, in which only artifacts (rather than full process traces) are typically available, and parallels live walkthrough methodologies commonly employed in HCI research \cite{mahatody2010state}.

\subsubsection{Evaluation of Second Junior's Artifacts (12 minutes)}
The procedure was repeated for a second junior participant's artifacts.

\subsubsection{Reflection (10 minutes)}
Seniors reflected on the utility of AI audit trails for mentorship and verification, as well as the role of AI tools in supporting their own learning processes.

\subsection{Data Ethics \& Analysis}

\begin{figure*}[t]
    \centering
    \includegraphics[width=1\textwidth]{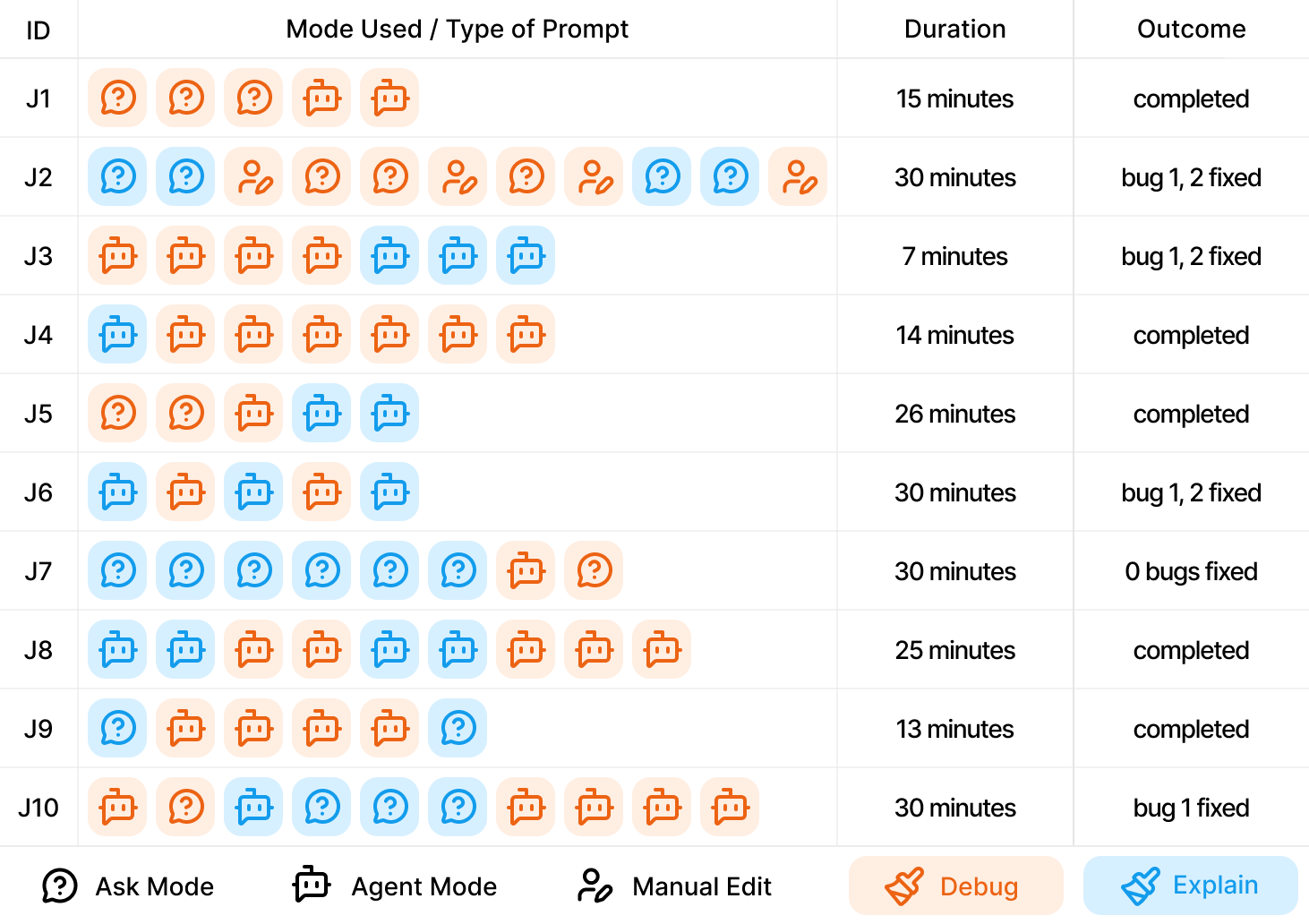}
    \caption{\textbf{Results from Phase 2 (P2).} Figure showing debugging task data for 10 junior engineers. Columns show participant ID from J1 to J10, detailed prompt sequences listing each interaction mode and type, task duration in minutes, and completion status. Prompt types include debugging, code explanation, codebase understanding, React concepts, and implementation suggestions. Task durations varied from 7 to 30 minutes with different completion levels.}
    \Description{Phase 2 junior engineers' debugging task workflows and outcomes. Comprehensive overview of 10 junior participants (J1-J10) showing their AI interaction patterns during the 30-minute debugging task. The table displays: participant ID, detailed prompt sequences with mode (Agent/Ask) and type (debugging / explanation), total task duration (ranging from 7-30 minutes), and final completion status. Results varied significantly: 5 participants fully completed all bugs, 3 completed the first and second bug, 1 completed only the first bug, and 1 did not fix any bugs at all. The prompt sequences reveal diverse approaches---some started with Agent mode for direct fixes, others began with Ask mode for understanding, and several switched between modes based on task progress.}
    \label{fig:P2_participants_prompts}
\end{figure*}

All procedures were approved by ETH Z\"urich's Ethics Committee (\texttt{Project 25 ETHICS-191}) and participants were informed of the risks and benefits of the study and signed an informed consent form prior to participating. Participants provided verbal and written consent for audio/video recording and for preserving prompt histories, code artifacts, and survey responses for analysis and subsequent, approved phases (e.g., artifact review). We stored data on encrypted drives, assigned pseudonyms (S1–S10, J1–J10), removed person- and company-identifying details, and deleted raw recordings after transcription and coding. Recruitment used convenience and snowball sampling (LinkedIn outreach, personal network and participant referrals). The team comprised one full-time industry software engineer (co-author) and two HCI researchers, all with computer science backgrounds. The industry author contributed domain realism in task design and interpretation of results. The interviews were analyzed via thematic analysis with grounded-theory–inspired open coding \cite{barke2023grounded}. 
Two coders (the same two interviewers that conducted all phase interviews together) independently coded all transcripts (including the tasks) from all phases, and their codes and content were then iteratively combined and organized into higher-level themes (the paper subsections) for reporting, with refinement based on discussion rather than numeric reliability. On-task artifacts from P1 and P3 were also analyzed with the codes, and artifacts from P2 (Cursor prompts, code diffs, postmortems) were directly reviewed by senior participants in P3, triangulating themes found from interviews. The Cursor prompts were also categorized as shown in Figure \ref{fig:P2_participants_prompts}, Table \ref{tab:agent_prompts} and Table \ref{tab:ask_prompts}.

\section{Results}

\subsection{RQ1: How do junior and senior software engineers allocate agency between themselves and agentic / generative AI in daily work?}

Here, we synthesize P1 and P3 seniors' boundary descriptions and P2 juniors' on-task decisions and descriptions of their daily practices. Our results give clear examples of who initiates changes when working with generative and agentic AI, when control returns to the human, and how engineers ensure they are answerable for their code.

\begin{figure*}[h]
    \centering
    \begin{subfigure}[b]{0.48\textwidth}
        \includegraphics[width=\textwidth]{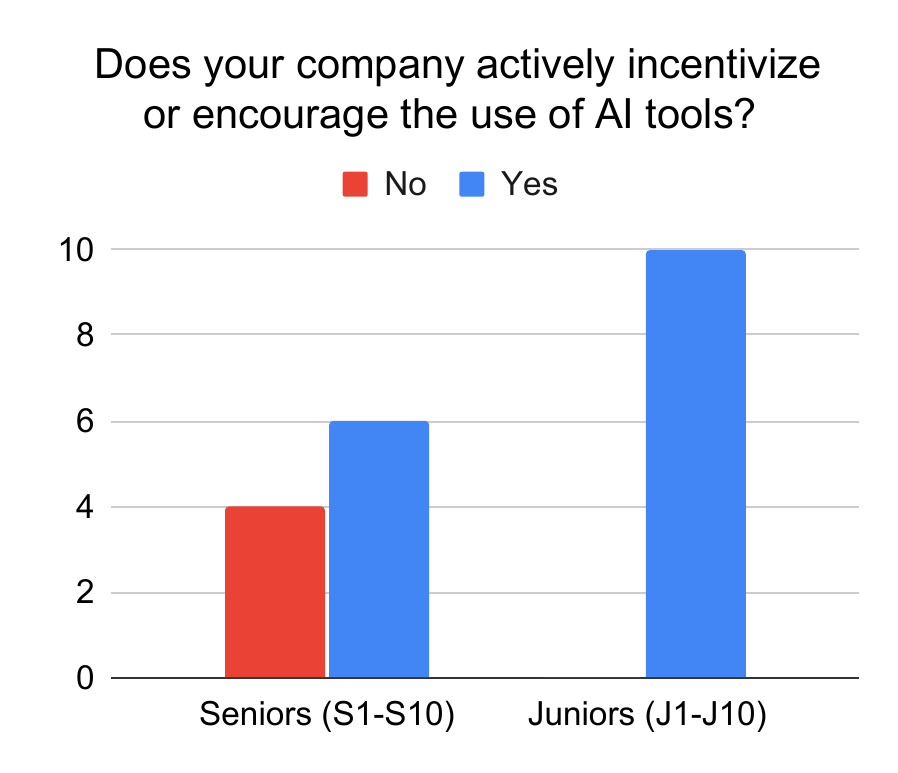}
        \caption{Company push}
        \Description{Question: "Does your company actively incentivize or encourage the use of AI tools?" 6 seniors said yes and all 10 juniors said Yes.}
        \label{fig:company_push}
    \end{subfigure}
    \hfill
    \begin{subfigure}[b]{0.48\textwidth}
        \includegraphics[width=\textwidth]{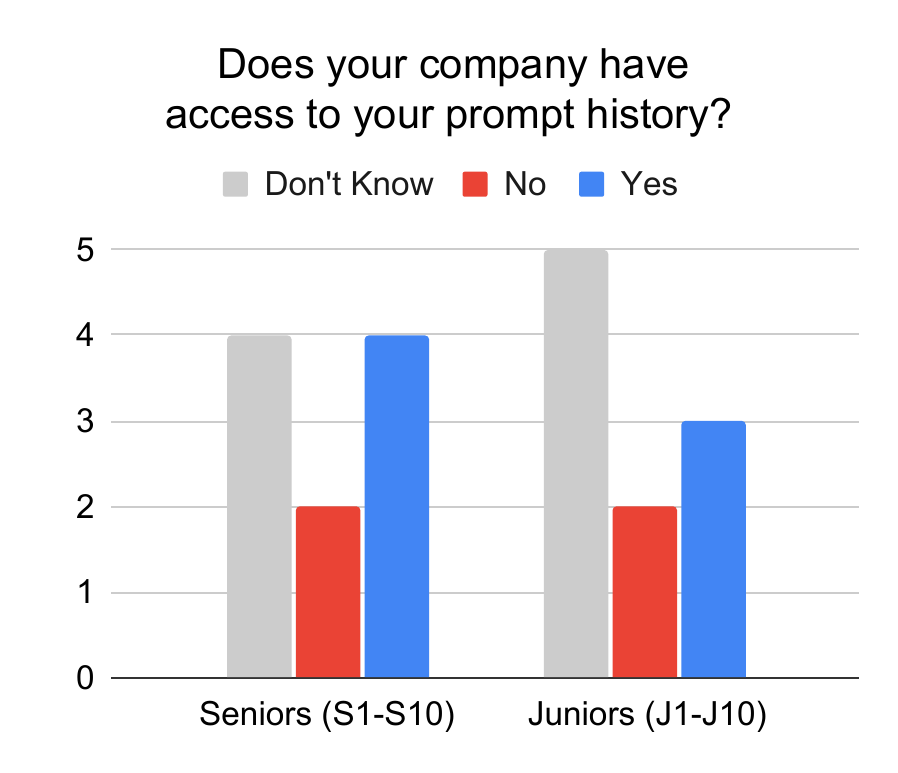}
        \caption{Access to prompts}
        \Description{Question: "Does your company have access to your prompt history?" For seniors, 4 answered Don’t Know, 2 said No, 4 said Yes; for juniors, 5 Don’t Know, 2 No, 3 Yes.}
        \label{fig:company_access}
    \end{subfigure}
    \hfill

    \begin{subfigure}[b]{0.32\textwidth}
        \includegraphics[width=\textwidth]{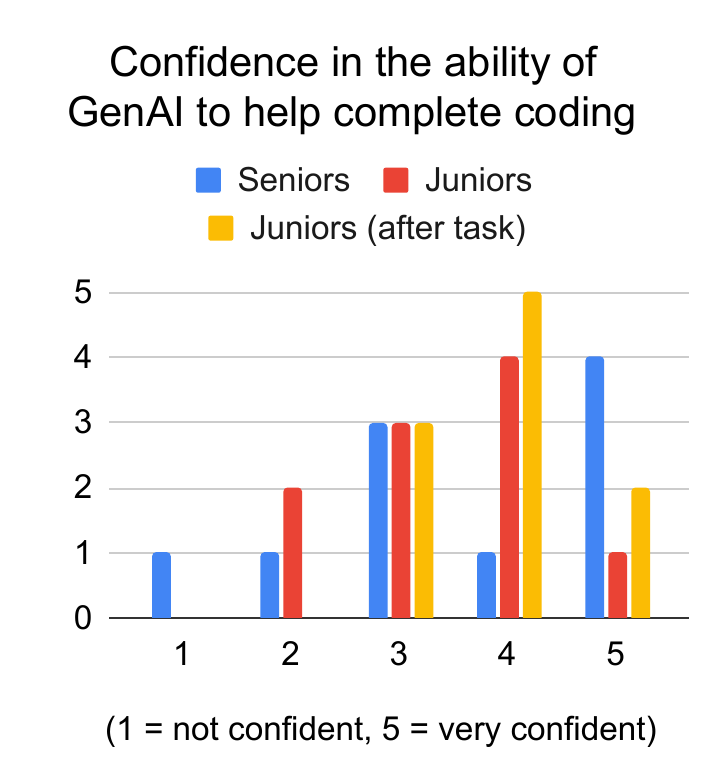}
        \caption{AI helpfulness}
        \Description{Question: "How confident are you in the ability of generative AI to help you complete coding tasks?" For seniors, 1 rated 1, 1 rated 2, 3 rated 3, 1 rated 4, and 4 rated 5. For juniors before their task, 2 rated 2, 3 rated 3, 4 rated 4, 1 rated 5. For juniors after their task, 3 rated 3, 5 rated 4, 2 rated 5.}
        \label{fig:ai_helpfulness}
    \end{subfigure}
    \hfill
    \begin{subfigure}[b]{0.32\textwidth}
        \includegraphics[width=\textwidth]{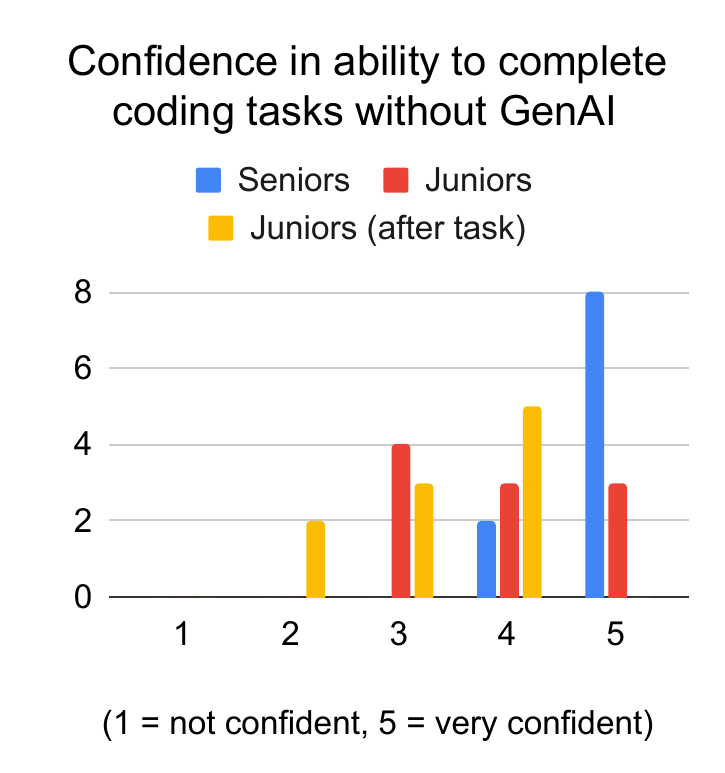}
        \caption{Confidence in self}
        \Description{Question: "How confident are you in your ability to complete coding tasks without generative AI?" For seniors, 2 rated 4, 8 rated 5. For juniors before their task, 4 rated 3, 3 rated 4, 3 rated 5. For juniors after their task, 2 rated 2, 3 rated 3, 5 rated 4.}
        \label{fig:confidence_self}
    \end{subfigure}
    \hfill
    \begin{subfigure}[b]{0.32\textwidth}
        \includegraphics[width=\textwidth]{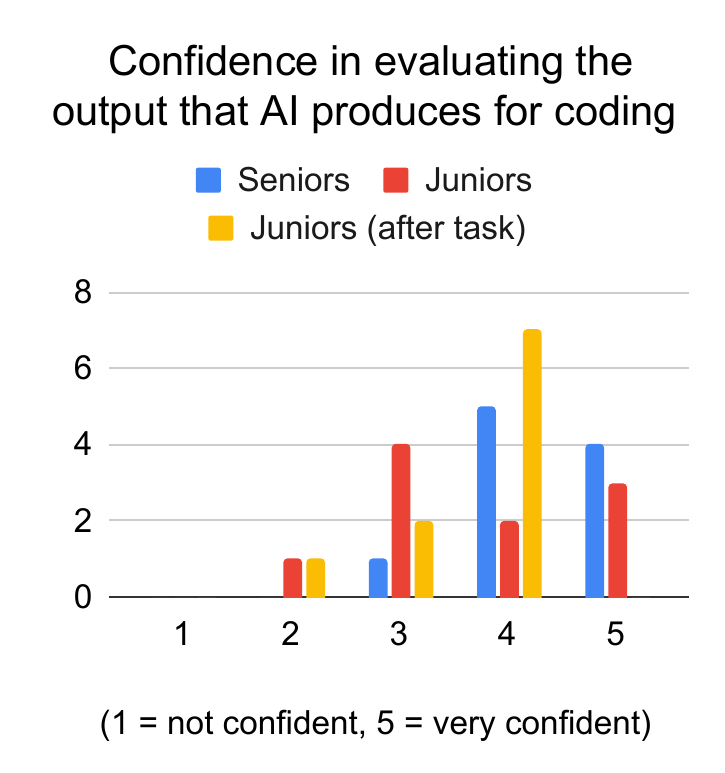}
        \caption{AI evaluation}
        \Description{Question: "How confident are you in evaluating the output that AI produces for coding tasks?" For seniors, 1 rated 3, 5 rated 4, and 4 rated 5. For juniors before their task, 1 rated 2, 4 rated 3, 2 rated 2, 3 rated 5. For juniors after their task, 1 rated 2, 2 rated 3, 7 rated 4.}
        \label{fig:ai_evaluation}
    \end{subfigure}

    \caption{\textbf{Survey responses across all three phases regarding AI adoption and confidence.} Comparative bar charts showing responses from Phase 1 seniors (n=5), Phase 2 juniors (n=10), and Phase 3 seniors (n=5). \textbf{(a) Company push}: 16 of 20 participants reported their companies actively incentivize AI tool usage. \textbf{(b) Access to prompts}: Mixed awareness about company access to prompt history, with plurality responding ``Don't Know'' (9/20). \textbf{(c) AI helpfulness}: Confidence in generative AI's ability to help with coding tasks, rated 1-5, showing moderate confidence across groups. \textbf{(d) Confidence in self}: Self-rated ability to complete coding tasks without AI, with seniors showing notably higher confidence (median 5) than juniors (median 3-4). \textbf{(e) AI evaluation}: Confidence in evaluating AI-generated code, where seniors unanimously rated 3-5 while juniors showed more variation (2-5).}
    \Description{Five subfigures showing survey data across three study phases. Figure a shows company AI incentivization with majority yes responses. Figure b displays prompt access awareness with most participants unsure. Figure c presents AI helpfulness ratings on 1 to 5 scale with varied distribution. Figure d shows self confidence ratings without AI where seniors rate higher than juniors. Figure e displays AI evaluation confidence with seniors rating higher while juniors show more variation.}
    \label{fig:survey_results}
\end{figure*}

\subsubsection{Company rules around AI}
\paragraph{\textbf{Company policies preconfigure AI agency boundaries before individual preferences matter.}}
Across phases, participants described mandates, allow-lists, and security constraints that preconfigure who writes what-an agent or a person-before individual tool preference ever enters. Companies varied in their rules for AI tool usage. Some had formal restrictions, such as limiting engineers to an approved internal set of tools with contracts in place (J8, J6, J4, S7, S3, S1, S10), prohibiting the use of non-company AI on corporate devices (S8), or requiring that confidential data not be shared with external services (J3, J10, J4, S6, S2, S1). In tightly regulated settings, participants named data to protect from AI-\emph{``Social Security numbers, credit card numbers, personally identifiable information''} (S2), echoing that their companies do not allow agentic AI and use old versions of Claude and ChatGPT (S2, S6, S10). Overall, AI tools were concrete and varied, as show in Table \ref{tab:P1}, \ref{tab:P2}, \ref{tab:P1b}. Companies also deployed internal knowledge bots, assistants that turned specs into code, Notebook LLMs, AI diagramming aids, and public prototyping tools like AI Studio (S8, J6, S7). In some cases, employees were required to use specific tools (e.g., Cursor) on a regular basis (J5, J9, S5), while in others there were no clear or memorable policies (J1, J2, S9, S4). Most companies actively pushed engineers to use AI (Figure \ref{fig:company_push}), while there was mixed knowledge on if their company has access to their prompt history (Figure \ref{fig:company_access}). Several (S8, S5, S4, S1, J9, J7) described a subtle loss of agency in the ``use AI now'' push from top-down, delivered matter-of-factly rather than as a choice. Within companies, seniors and juniors mentioned various capabilities that were tacit to their role that limited their use of AI, as often AI can not reliably infer infrastructure conventions, tooling, and history. S3 offered an example: AI that proposes a virtualenv-based training pipeline that \emph{``runs''} but cannot be checked in due to incompatibilities with company specific Bazel and Pytorch versions. S1 added that a product's \emph{``true nature''} often lives in distributed mental models; their team's mitigation is to externalize key knowledge so progress is not person-dependent. Similarly, participants (J9, J8, J7, J6, S9, S2, S10, S8, S4) raised concerns about answerability for AI-generated code, with J9 noting, \emph{``there's a lot of code ... [seniors] have to trust that you tested ... sometimes you wake up and something is broken''}. 

Overall, seniors reported high self-confidence in coding without AI, more varied views on AI's usefulness, and strong confidence in evaluating AI output (Figure \ref{fig:confidence_self}, Figure \ref{fig:ai_helpfulness}, Figure \ref{fig:ai_evaluation}). After completing the coding task with agentic AI, juniors demonstrated varied responses in their self-assessed confidence for ``Confidence in ability to code without AI'' (Pre-Task $M = 3.90 \pm 0.88$) to Post-Task $M = 3.30 \pm 0.82$), Cohen's $d = 0.71$). The majority (6 of 10 participants) reported decreased confidence post-task, with changes ranging from $-1$ to $-2$ points on the 5-point scale. This trend was particularly notable among those who initially rated themselves highest (at level 5): all three of these participants decreased their ratings post-task. In contrast, self-confidence in evaluating AI output had diverse trajectories pre and post task: 3 participants decreased their ratings, 4 increased, and 3 remained stable. Notably, post-task ratings converged substantially toward level 4, with 7 of 10 participants (70\%) rating themselves at this level, compared to only 2 participants (20\%) pre-task. This convergence suggests that the task experience may have calibrated participants' meta-cognitive awareness of AI evaluation, standardizing initially varied perspectives (Figure \ref{fig:confidence_self}, Figure \ref{fig:ai_helpfulness}, Figure \ref{fig:ai_evaluation}).

\subsubsection{Agency in cases of high familiarity}

\paragraph{\textbf{Seniors maintain control through detailed delegation and iterative refinement.}}
When tasks, context, and code were well-understood, seniors stayed at the wheel: they scoped the change, explaining to AI in great detail, and delegated small units to AI (syntax, boilerplate, tests, formulas), insisting on minimal, reviewable diffs (S10, S4, S8, S1, S5, S7, S3). Seniors differed on the tasks they were comfortable with doing with the AI; it ranged from basic tasks or tasks that do not require much context (S1, S5, S10, S7) to taking a design and translating it into code by going back and forth with the AI (S4, S8). Refactors with many dependent codebases were brittle and required a lot of detail, with S1 stating that \emph{``any refactoring task which is straightforward for a human \dots you will find that [AI] struggles a bit \dots make your prompt as comprehensive as possible to explain what you're trying to do''} (S1). Seniors (S10, S3) guarded against model assumptions about ``typical'' codebases (rejecting YAML/JSON edits that misfit local conventions) and trimmed bloat: \emph{``AI usually generates too much CSS that you don't need.''} (S2). Their stance (S4, S10) was designer-like: arrive with a plan, delegate only what fits, and iterate until intent is met: \emph{``I'm usually already going in with the idea \dots it might not survive implementation, that's fine.''} (S10). Seniors also seeded context-e.g., feeding exemplar code links and asking the model to match patterns (S1)-and leaned on joy-based delegation: \emph{``If you don't like writing tests, throw that to the AI \dots if you think CSS is fun, then keep doing that''} (S2).

\paragraph{\textbf{Juniors use constraint-based approaches with heavy verification and scope limiting.}}
With frameworks and codebases they already knew, juniors similarly constrained AI to small, targeted edits and recall: \emph{``I wouldn't use an AI to do something that I don't think I could actually do.''} (J6). They broke work into sub-steps to be \emph{``absolutely sure of what's going into your code before you approve it''} (J9), scoped files to change first (J4), or simply coded the known fix by themselves (J4, J1, J5, J8 when the bug was clear, J3 when the AI output seemed wrong). Several also described a design-led style: on a single service they know deeply, write precise prompts then manually test, contrasting this with broader, heavier agent use (J9, J4). For debugging in familiar repositories, they pasted errors/stack traces (J2, J10, J8, J9) but avoided agents when diffs might sprawl beyond reviewable scope (J6), echoing the sentiment that agents such as Cursor output too many changes and information at once (J7, J6, J5, J10, J9, J8). For ``boring''/low-stakes tasks (tests, renames, formatting, syntax, sql queries, documentation), AI saved time but sometimes still needed correction (J6, J8, J4, J9, J2, J7, J3, J1). Cross-checking (Google, a second model, coworkers, exemplar code) was routine (J1, J5, J8, J4). In addition, J6 and J7 noted that before they use AI, they have a vision of what is right and wrong and what they think the code or answer will be; in familiar codebases they would trust themselves over the agent (J3). Similarly, for the P2 task, since J1 already knew React, they stated they knew the places to check and how to prompt; J9 also stated that they knew exactly what to prompt since they had used Cursor for React once before. J8 stressed the ability to read through the suggestions and modify them, especially when the suggestion is wrong/needs additional detail; in P2, Cursor sometimes made numerous irrelevant edits: \emph{``it started spiraling \dots I just stopped it. The fix was a three-line change''} (J6), whereas J1 caught subtle missteps (e.g., an unnecessary timestamp param) and J10 stopped the AI from generating unneeded mock-data.

\subsubsection{Agency in cases of low familiarity}

\paragraph{\textbf{Seniors separate design from code generation, maintaining strategic oversight.}}
In unfamiliar areas, seniors used several sensemaking techniques with AI: understanding how the code maps to interactions, asking for pros and cons on a specific design, generating diagrams/overviews to internalize structure, generating a basic template, and asking for ideas even if they have their own ideas (S8, S10, S1, S5, S4). S6 stated that they would prompt \emph{``how would you start debugging this? Is there some better approach I should be using,''} while S1 noted that they had to wade through irrelevant ideas: \emph{``I did end up feeding [the bug] to a chatbot ... it did generate a list of 4 or 5 insights. Some of them were stylistic comments like, where you can fix this, you can do this better and do this optimization and hidden in that list was one of those insights that you're doing this wrong.}'' Some favored questions that can't easily mislead, such as \emph{``can you show me the code path''} (S4), only using AI to confirm their expectations about how the code works rather than asking AI open-endedly (S4, S7, S10). Others using agents only for small straightforward snippets (e.g. generating JWTs), or for library syntax they didn't recall (S10, S9 S5, S3). They also used AI as a template to build upon (S4, S10, S9), and a recurring guardrail prompt was to addtests that indicate that changes are harmless (S2, S6, S10). Regardless of familiarity, seniors emphasized that engineers still need to have the common sense on how to drive the project later, even with AI help.

\paragraph{\textbf{Juniors oscillate between over-reliance and defensive resistance, struggling with scope control.}}
Without a robust mental model (new repository and or unfamiliar with React), agents widened scope. For the P2 task, J8 and J2 stated that the new codebase made it harder to figure out where the issues were happening, so for them the AI was very helpful in terms of parsing all the files in the codebase in a short amount of time. Though participants (J4, J10, J8) tried to retain control by shrinking edits, over-reliance surfaced; J3 stated that for unfamiliar codebases, they trust Cursor more than they trust themself until they gain context through peer review, while J10 noted that \emph{``when you're coding with agents, it's like you're just free-falling \dots I could see myself easily pressing \dots spamming the agent button at my work.''} J2 noted that they always ask AI first before doing any debugging, and outages exposed dependence-\emph{``Copilot was down \dots I feel like I've become a lot more reliant on AI''} (J2). Juniors also emphasized manual testing (J9, J8, J5), even if they do not completely understand the AI's changes (J2, J3). During the P2 task, some switched to Agent mode because Ask mode \emph{``assumes the app works''} (J9), or because Cursor \emph{``seems like it's known this codebase forever''} (J10). 

Several contrasted ``what I'd do with more time'' (read broadly and do more surgical operations) with ``what I did here'' (lean on the agent and then back-fill understanding) (J10, J7, J2). However, others still tried to regain control, and used Ask mode for open ended questions to confirm high-level suspicions or fill in React concepts (J6, J5, J7). Others watched closely and intervened on correct or incorrect gut checks-declining broad deletions to fix a narrow bug (J4) or finding it wrong that the Agent mode commented out the service worker (even though it was not wrong) (J10). Some who let the agent draft first corroborated results and took notes for learning (J5, J8). Others rejected all agent suggestions and requested a minimal refactor, refusing to read giant diffs (J4, J7, J8). Some avoided the use of AI altogether, preferring to ask other engineers (J7, J5), or using Reddit, Stack Overflow, Google or even other LLMs (J5, J4). Interestingly enough, for P2, participants with limited React knowledge had different approaches, while J10 heavily relied on Agent mode, J7 stated that they should have relied on the agent mode a bit more, though they were hesitant because they were unfamiliar with how Javascript worked and asked conceptual questions instead. P2 additionally highlighted how time shapes control, with many stating that they would not use AI if they had a lot more time (J2, J7, J5, J9); some who preferred Ask mode in Cursor switched to Agent mode (J5, J7).

\subsubsection{AI use during task completions}
\paragraph{\textbf{Seniors view AI as support, not a replacement for system understanding.}}

During the ACTA process, seniors observed that for debugging, AI was only effective when fine-tuned to the specific service or provided with sufficient prompt context; otherwise, it risked generating hallucinated outputs (S1, S5). For feature creation, S2 and S4 found AI most useful for code generation, while S3 highlighted its value for data manipulation. However, S2, S3, and S4 emphasized that AI should not replace the engineer's understanding of the system or the code produced. See Appendix~\ref{sec:acta-brainstorm} for each of the five senior's fully brainstormed task-completion steps.

\paragraph{\textbf{Juniors held diverse patterns in problem-solving initiation and code comprehension.}}

\begin{table*}
    \centering
    \caption{Table displaying the types and frequencies of prompts used in Agent mode during Phase 2. Categories include debugging, code explanation, command, concept explanation, and asking for ideas. Debugging prompts show the highest frequency, followed by code explanation prompts.}
    \Description{Distribution of Agent mode prompt types used by junior engineers during debugging task. Table showing the frequency and categories of prompts submitted when participants used Cursor's Agent mode (autonomous code completion). Prompt types are categorized by purpose: debugging, code explanation, command, command/code explanation, command/debugging, concept explanation, and asking for ideas. The distribution reveals that debugging prompts were most common, followed by requests for code explanation.}

\label{tab:agent_prompts}
\renewcommand{\arraystretch}{1.5}
\begin{tabular*}{\textwidth}{l p{1.5in} p{0.3in} p{4.35in}}
    \toprule
    Mode & Type & Count & Sample Prompt \\
    \midrule
        Agent & Debugging & 24 & When navigating to User Logs and entering a user ID, the app stalls on Loading Activity logs \\
        Agent & Code explanation & 8 & where is the div card-actions being used in the code \\
        Agent & Command & 4 & I don't want to remove the existing filter and pagination logic \\
        Agent & Command/Code explanation & 2 & look at the rest of the UserLogs file. Is there anything that modifies userInput? \\
        Agent & Command/Debugging & 2 & it seems like the "user" object is not getting set. find out if there is something wrong with either the "setUser" part or elsewhere \\
        Agent & Concept Explanation & 2 & what does "window.location.href" do \\
        Agent & Asking for ideas & 1 & how would you test the handler to ensure that it does not contain any URL routing logic? \\
        \bottomrule
    \end{tabular*}
\end{table*}

\begin{table*}
    \centering
    \caption{Table displaying the types and frequencies of prompts used in Ask mode during Phase 2. Categories include code explanation, debugging, React concepts, and verification. Code explanation show the highest frequencies, contrasting with Agent mode usage patterns.}
    \Description{Distribution of Ask mode prompt types used by junior engineers during debugging task. Table showing the frequency and categories of prompts submitted when participants used Cursor's Ask mode. Prompt types are categorized by purpose: code explanation, debugging, concept explanation, command, and verification. The distribution reveals that code explanation and debugging were most common, followed by requests for concept explanation.}
\label{tab:ask_prompts}
\renewcommand{\arraystretch}{1.35}
\begin{tabular*}{\textwidth}{l p{1.5in} p{0.3in} p{4.37in}}
    \toprule
    Mode & Type & Count & Sample Prompt \\
    \midrule
        Ask & Code explanation & 12 & What is the logs-container div doing? \\
        Ask & Debugging & 10 & Are there any other changes that might need to happen? The issue is still occuring \\
        Ask & Concept Explanation & 3 & When does useEffect() run? \\
        Ask & Command & 1 & generate a postmortem on this bug: cause of issue, how to debug, how to prevent it happening in the future \\
        Ask & Verification & 1 & Is what I have commented out correct? \\
        \bottomrule
    \end{tabular*}
\end{table*}

\begin{figure*}
    \centering
    \includegraphics[width=0.7\textwidth]{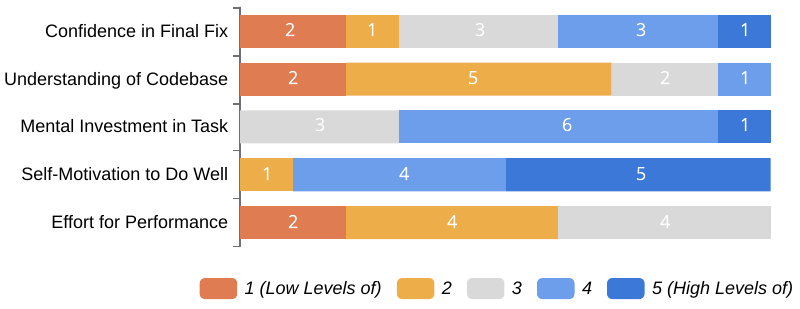}

    \caption{\textbf{Post-task survey responses from Phase 2 junior engineers.} Five bar charts displaying participants' self-reported assessments immediately after completing the 30-minute debugging task. \textbf{Confidence in fix} (1-5 scale): Most rated 3-4, indicating moderate confidence in their solutions despite varying completion rates. \textbf{Current codebase understanding} (1-5 scale): Half selected 2, with none rating 5, reflecting limited comprehension of the unfamiliar React application after the brief task. \textbf{Mental investment} (1-5 scale): Predominantly 4 ratings indicate high cognitive engagement despite AI assistance. \textbf{Importance to do well} (1-5 scale): Most rated 4-5, demonstrating strong motivation to succeed in the research task. \textbf{Hard work required} (1-3 scale): Most rated 2-3, suggesting the task demanded significant effort. These results highlight that while juniors remained engaged and motivated, they struggled with confidence and understanding when debugging unfamiliar codebases, even with agentic AI support.}
    \Description{Five bar charts showing post-task survey responses from 10 junior engineers on a 1-5, with 1 representing "low levels of..." and 5 representing "high levels of", so for the first question a 1 would be "low levels of confidence in final fix". Charts display confidence in fix with most rating 3 to 4, codebase understanding with majority at 2, mental investment predominantly at 4, importance to do well mostly at 4 to 5, and hard work mostly at 2 to 3. Response distributions indicate moderate confidence but limited understanding despite high effort and motivation.}
    \label{fig:post_task_survey_results}
\end{figure*}

After completing the task, participants' confidence in their solutions varied, and most reported limited understanding of the task codebase and lower effort in completing the task, as shown in Figure \ref{fig:post_task_survey_results}. Despite this, most juniors remained mentally engaged and motivated to perform well. Notably, some juniors adjusted their self-ratings, reporting slightly lower confidence in completing tasks without generative AI, while expressing increased confidence when using generative AI.

Most of the juniors started out by using Ask mode (J2, J5, J7, J9) or Agent Mode (J4, J6, J8) to understand the codebase and how specific functions worked, while others directly jumped into solving the issue with Agent mode (J1, J3, J10). Some of the juniors inputted their own opinions and suggestions into Agent mode and asked for feedback and ideas (J1, J2, J6, J8), others asked for explanation on React concepts not specific to the codebase (J7, J8), and others just described the issue directly (J1, J4, J5, J10) while many asked questions without inputting their opinions (J1, J3, J9, J4, J2, J7, J10). J5 went back after using Agent mode to solve all three bugs and asked clarifying questions about how the fix worked, taking notes for their own learning. Overall, a majority of prompts that were used with Agent mode fell under debugging, while Ask mode showed more of a diversity with code explanation and debugging prompts.

\subsection{RQ2: How do junior and senior software engineers perceive professional growth for a junior in the age of AI?}

We triangulate P1, P2, and P3 retrospective accounts of their growth, as well as their methods for learning with AI.

\subsubsection{Career growth (during or before) the age of AI}

\paragraph{\textbf{Seniors developed foundational instincts pre-AI that remain critical for steering modern tools.}}
Seniors reflected on pre-AI tools and identified critical practices from those experiences that remain essential when working with AI. S3 contrasted tools they had in the beginning of their career with today's tools: \emph{``The closest thing you had to AI autocomplete was just IntelliSense  \dots  I don't think language servers were even a thing really when I first started \dots now with AI, I think things have changed dramatically.''} S9 noted using various resources (e.g., Udemy, Coursera, and Youtube) to learn explicit knowledge, while they learned from senior engineers by looking at how they code and what tools they use.  S2 and S5 admitted early-career overconfidence-wanting to own everything and build complexity-tempered by failure. All the seniors noted practices that helped them grow; S2 stated \emph{``you have to take down the database in production so you know not to do it''}, while S7 emphasized reading lots of code-not only their own code-to develop their skills, stating that it builds a mental map: \emph{``[you] get an intuition of how all these pieces connect \dots a mental map that you can go and say I know either who to reach out to or I know where this thing might be.''} S6 and S2 echoed the mental map-based fluency, noting how seniors often have to hold a lot of complexity and conditions in their mind. General software engineering instincts also developed across time; S3 flagged scoping failures and insufficient observability as red flags you feel, stating \emph{``you can kind of tell when it feels off''}, while S2 noted that \emph{``I can just look at something \dots and go `this isn't gonna work very well'.''}

Seniors (S6, S10, S7) framed AI as one tool among many on a junior's path to becoming senior and many (S5, S7, S1, S9, S2, S3) emphasized not reinventing the wheel and using existing libraries and code, even when generating with AI. S6 placed AI alongside editor, browser, and headphones: always there, but never in charge. Similar to how S2 and S9 emphasized learning from trial and error, seniors endorsed AI as a \emph{``faster, more interactive Google \dots and a rubber duck''} and even a way for juniors to \emph{``create messes faster''} so they can learn from them (S6). Similarly, S3 claimed juniors can now ramp on build systems faster by asking agents instead of struggling blindly. However, S10 and S4 warned that growth still requires building one's own intuition, with S10 advising that \emph{``the AI's not handling the high level problem solving.''} S7 also observed fewer engineers \emph{``offering up a theory''} of what might be happening; they fear eroded code reading and critical thinking skills in juniors: \emph{``I feel like junior engineers just have no intuition. Like, I am routinely pinged on stuff like, hey, this thing doesn't work and with zero follow-ups on a speculation.''} S7 also commented that juniors might eventually rely too heavily on seniors: \emph{``they're not necessarily thinking through the potential consequences of the code they're writing \dots it ends up being a senior engineer, inevitably, who's like, oh wait, did we do something like this?''} S8 was blunt saying that much of junior work \emph{``can be replaced with AI''}, and that juniors will simply need to learn faster. S7 and S4 note that the most successful engineers are those who read AI-generated code, think through why it works, and integrate it thoughtfully, rather than just rushing to complete a task.

\paragraph{\textbf{Juniors achieve speed gains but struggle with ownership and imposter syndrome in AI-native careers.}}
Juniors held different ideas of productivity with AI, with overarching themes of feeling loss in ownership and fulfillment in their work. Many juniors have never known professional software work without AI, with J4 stating \emph{``the whole time I've been \dots a software engineer, there has been AI \dots I don't really have a point of comparison.''} Juniors credited AI with freeing time and enabling parallelization. J2 noted that speeding delivery increases breadth and thus learning: \emph{``[you] get to work on more things in a shorter amount of time.''} J8 and J1 agreed, stating the extra time allows them to \emph{``talk to coworkers more about like design tasks, or more tasks that require a little more human input opinion like when tasks where there's no right answer.''} (J8). AI tools also enabled more efficient debugging, as engineers could paste in large error logs (J2, J3) that Google could not handle and that Stack Overflow lacked the context to address. However, leaning on AI tended to cause imposter syndrome. J3, J7, and J8 noted that if they use AI to author code without understanding, they would feel \emph{``ill at ease''}, with J8 explaining, \emph{``It has my name on it, but I have no idea why it works''}. Both J4 and J9 still felt accomplishment when they were involved in the design process, but when even design was delegated, J8 noted that it \emph{``doesn't feel good when you finish something really quickly \dots you're like, I just clicked on AI.''} After being praised for speed during their task, J3 demurred: \emph{``Cursor did the work \dots I just tried to find the problem.''} Similarly, J9 stated using Cursor at a prior hackathon made them feel \emph{``like a fraud.''} J2 brought up how being told the answers from AI hinders feelings of personal achievement, explaining \emph{``it's kind of like if someone gave you like all the answers to test.''}

\subsubsection{What makes a senior ... senior?}

\paragraph{\textbf{Seniors distinguish experience levels through system-thinking and predict AI will elevate non-technical skills.}}
S5 noted that senior diffs tend to be smaller and safer: \emph{``the smaller your changes are, the less likely it's going to lead to issues.''} S1 noted that juniors own low-ambiguity refactors and seniors reshape systems. S2 also noted the explicit skills learned on the job, lamenting that in school, \emph{``they always told us `test your stuff' but they never told us how.''} Similarly, S3 and S10 flagged an anti-pattern in junior PRs: trying to ship large amounts of changes in one pull request, with S3 stating that \emph{``coding is really about \dots trying to make things simple for everyone.''} Several seniors (S2, S3, S5, S1, S4) emphasized big-picture thinking: how one change could affect another, understanding how all the elements fit to identify problematic architecture, and to avoid drilling too deep into a specific area (i.e. being led astray). Despite the difference in title, two seniors cautioned about making too much of a distinction between juniors and seniors. S9 emphasized that titles matter less than \emph{``smartness and passion''}, while S1 articulated level boundaries: juniors deliver clear tasks independently; mid-level engineers handle complexity and guide others; seniors solve complex problems with simpler solutions. Finally, S10 noted that some seniors could actually lack best practices when steering LLMs compared to AI-native junior engineers who started their careers with AI.

On how AI may shifts senior's expectations of juniors, opinions were varied. Many seniors foresaw a sense of elitism around being able to code without AI and identifying AI generated code. S3 also insisted on the importance of baseline competence without the help of AI and the ability to communicate work clearly to others. S8 noted the expectation that new hires grasp systems faster with AI---even without knowing the deep internals---and that they understand high level architecture and interactions. S5 further anticipated continued abstraction creep, similar to how \emph{``assembly or C++''} is abstracted away. Finally, seniors anchored growth in non-code work, with S1 stating, \emph{``coding is the easiest part''} and that seniors \emph{``set the direction for the team''} and help leadership allocate time (S1, S2). S4 emphasized learning to identify \emph{``who is gonna be your champions and who are going to be people in your way.''} Overall, seniors emphasized the development of ``soft-skills'' over ``hard-skills'' in the age of AI.

\subsubsection{AI for learning}

\paragraph{\textbf{Seniors note that AI accelerates learning but cannot replace understanding.}}
Seniors acknowledged the usefulness of AI, comparing it to existing tools. S4 treats AI as an always-on mentor: \emph{``anytime I need to understand something, I just go back and forth with Claude until Claude agrees that I have a correct understanding''} S9 also noted that junior engineers can learn new technology from AI more efficiently and easily, and S6 noted that they use AI to generate a lot of \emph{``really small, basic examples.''} However, S2 warned against trusting AI without critical thinking: \emph{``don't take it at face value. Do your research,''} and S7 and S2 further explained that use of AI is as dangerous as using Stack Overflow without thorough understanding, where they used to ask juniors \emph{``did you read the whole answer before you copied the code...we didn't do that. We won't do that now either [with AI].''} S6 noted the importance of gaining experience through consequences: \emph{``err on the side of creating messes because you learn from creating those messes.''} Most importantly, S4 and S2 cautioned juniors not to let AI displace understanding, with S4 stating, \emph{``at no point can you hand over your expertise. You're just handing over the workload.''}

\paragraph{\textbf{Juniors navigate tension between learning efficiency and skill atrophy, with mixed results.}}
Several (J2, J5, J6, J9) said they \emph{``learn by doing''} and explained that simply copy-pasting or generating code is not very helpful. J1 shifted from YouTube crash courses to prompting AI for \emph{``a really small example that I can play around with''}, praising the added interactivity. J8 and J9 stated that they would not have the necessary knowledge to know if something is best practice, so they would ask the AI to explain its reasoning for ``why it chose something'', taking the the answers if they feel logical.Several juniors also mentioned avoiding the use of AI altogether, citing tech-debt as a major concern (J3, J7, J10). J7 stated, \emph{``especially as a someone towards the start of my career \dots this is where I need to just learn as many things as possible \dots If I go, and I ask a AI agent to do it, I might be faster this time but I might not be faster next time.''} Several juniors (J1, J3, J8, J9) cited worries about their ``critical thinking muscle'' atrophying due to overreliance on AI. Indeed, when J3 was asked what they normally do while the agent buffers, they said, \emph{``what I do is go to Instagram,''} explaining a potentially dangerous trend of full context-switch and over-reliance on tools with the availability of agents.

\subsection{RQ3: When and why do engineers deem mentorship indispensable in workflows with agentic / generative AI?}

All senior engineers noted that they mentored either junior engineers, teammates, or interns during their career either through code review, design conversations, and debugging, and that they do code review almost every day.

\subsubsection{When AI can be a mentor}

\paragraph{\textbf{Seniors position AI as an accessible fallback mentor for basic guidance.}}
Seniors emphasized the usefulness of AI when there are no other mentors available. S7 underscored that AI can surface best practices, noting that \emph{``[AI] can help show them some patterns which [juniors] are not aware of.''} In terms of having AI as a mentor, S4 mentioned that at their previous company, they got the criticism that they would bug senior engineers too much and that AI replaced much of the question-answering from seniors. S6 also stated that AI is a good alternative when seniors are not around. J10 also recalled a time where they had an engineer mentoring them and at the end the engineer said if they have any other questions they can ask GPT, reflecting, \emph{```He trusts ChatGPT to teach me the same way that he would teach me.''}

\paragraph{\textbf{Juniors embrace AI as an always-available first mentor to reduce dependency on busy seniors.}}
Juniors used AI as a mentor for basic questions that required little to no context. J2 likened AI chats to \emph{``talking to \dots a smart person that knows everything''} and admitted they ping teammates less. J9 stated that most of the time their senior engineers are too busy so they cannot do thorough code review. Explaining that while \emph{``[Agentic AI] does reduce the number of times I have to ask,''} it changes expectations of when one goes to seniors for help, as when they used to be an intern at a company, their mentor implictly suggested that \emph{``there's an answer for everything on the Internet''} and that J9 took it as a \emph{``why are you asking me so many questions?''} J1 delegates simpler questions to AI, such as \emph{``how do I find this in a file [or] what does this function do?''} while J8 states that their job requires a lot of finance knowledge, which they did not learn in school, so AI was very helpful for teaching them.

\subsubsection{When AI cannot be a mentor}

\paragraph{\textbf{Seniors act as critical guardrails against AI's context blindness and overconfidence.}}
S1 and S6 emphasized the senior's roles in stewardship and unlocking a junior's potential. Similarly, S9 highlighted the uniquely human, organization-specific guidance seniors provide. S10 described mentorship-by-questioning to instill a sense ownership: they would ask juniors why they chose an approach and what they expect to see next, arguing that today's models tend to agree uncritically and contribute to loss of critical thinking. 

Several seniors (S1, S3, S4, S5, S6, S10) mentioned that AI can confidently lead a junior to a solution that they are not (or should not be) looking for. Thus, seniors also act as guardrails against hallucinations and bad practices when they review a junior's code output. S2 had to correct agents repeatedly on pinned versions (i.e. \emph{``Next.js 15 versus 14''}), and S3 warned that agents will happily \emph{``relax the type''} or \emph{``ignore the compiler''} to satisfy a command like \emph{``fix this type error,''} which a novice programmer might simply accept. S10 noted that juniors try to point AI at a problem, accept the output if it passes tests, and move on.While this may speed up progress, it's short-sighted because it ignores longer-term concerns like maintainability and code quality. S6 further commented on how they mitigate tech debt, explaining that when they conduct code reviews, they pay attention to red flags that indicate that the junior is not sure what they are generating, such as \emph{``building things we don't need right now, you can just tell, nowadays ... it's probably AI.''} S2 describes his approach to using code reviews for mentoring and knowledge transfer: \emph{``I try and give concise comments on code and try and also point to documentation and \dots link to a Stack Overflow answer that I found useful.''} While seniors can act as guardrails, S4 also flagged code review fatigue: AI can generate \emph{``tons of code''} and \emph{``it can be beyond the point where a senior engineer might implicitly trust a junior engineer and and let code go out that shouldn't or they get exhausted from code reviewing, and they're not able to grok everything \dots in that code review.''}

\paragraph{\textbf{Juniors recognize AI's limitations and seek human insight for context and complex decisions.}}
Juniors noted that consulting humans is still crucial for clearing up understanding and discussing context-specific topics. J7 reaches for humans for /emph{``why-questions''}: \emph{``when the reason for a decision is sitting in someone's head, I can't expect to get that answer from an AI.''} Similarly, J4 said the original author of a piece of code should be able to \emph{``explain it \dots way better and faster [than AI].''} J6 also noted that a lot of their work is context specific, and that sometimes their LLMs that are trained on their company's sparse documentation hallucinates and leads them astray, even when it is the AI is good at citing its resources and adding links. For code review, some juniors stated that even though they use an AI code reviewer, it \emph{``doesn't catch things, and in terms of more complex design, it will turn up a lot of false positives''} (J9). Similarly, J6 once experimented with an AI code review tool but found its feedback overly verbose and unhelpful. Finally, J7 also mentioned that they enjoy talking to people more than they like reading a big wall of text that the AI has generated.
\subsection{RQ4: How do agentic AI records (e.g., prompt history, provenance) shape code review and mentorship?}

\subsubsection{Prompt history review feedback.}
Reviewing specific traces from P2, seniors suggested the following
\begin{itemize}
\item \textbf{Start with understanding.} S7 favored J4's Ask mode prompts to first to understand what was happening. S10 also noted they would have Cursor \emph{``summarize the codebase and the high-level issue} unlike J6. S8 wanted juniors to be able to explain and defend an agent's proposed change; when J10 kept prompting without reading, S8 saw a \emph{``prompting spiral''}: \emph{``after the fifth try, the issue's still not fixed.''} S10 also emphasized asking the AI for a summary of debugging tactics, and S4 similarly emphasized the importance of asking the AI questions.

\item \textbf{Tighten prompts.} S8 showed J9 could have merged two early prompts (\emph{``What does load logs \dots  point to''} + \emph{``When Load Logs is clicked, redirects to main dashboard''}) into a crisp bug description. S9 wanted more detail in J3 and J5's prompts, and also noted that context for Cursor should be meaningful, rather than the one line of code that J3 had pasted in.

\item \textbf{Prefer small, testable batches.} S10 praised J8's small prompts, S6 emphasized small-batch development, and S7 liked J4's Agent mode sequence that yielded a clean, reviewable diff, compared to commented out code in J2's Ask mode sequence. S10 and S6 wanted tests accompanying J6/J8/J1's changes.

\item \textbf{Proper use of Cursor modes as well as basic tools.} S7 agreed with J2's start of using Ask mode to look at the code to explain what might be happening: \emph{``they're articulating that \dots  the behavior that they expect to see isn't happening and leveraging cursor to kind of inspect the code more broadly and identify what's wrong with it, which is good.''} S9 also agreed with the switch from Ask mode to Agent mode for J5 for making code changes.  Across five senior reviews, three (S9, S10, S7) pushed juniors to pair AI with core tools-grep, debuggers, logs-rather than use AI as a first and only resort. S10 called some traces \emph{``a kind of a lazy use of AI''} and reminded juniors to \emph{``know the tools available to you.''}

\item \textbf{Constant vigilance.} S6 was happy that J1 was able to call out the AI on the fact that its suggestion did not work; in contrast, S7 mentioned that J4 should read the diffs meticulously to see if AI changed anything unexpected, as their code review showed additional changes not related to fixing the bugs. Similarly, S10 noted that juniors should think about changes beyond what the AI makes; if AI changes one usage pattern (e.g., React navigation), S10 wanted that consistency across the repo, not one-off patches. S6 also noted that \emph{``some of these prompts look like their cognitive load's being reached, and they're compensating and asking for the AI to help \dots as long as the AI doesn't lead them astray, I think it's great.''}
\end{itemize}

\paragraph{\textbf{Seniors infer junior competence and thinking patterns from prompt quality and specificity.}}
After reviewing the junior's prompts from P2, seniors noted several feedback and words of wisdom they wanted to tell the junior. S10 stated that they would ask J6 where they got the idea of prompting \emph{``how do i persist isloading across dom refreshes,''} to understand their thought process. S9 noted that Cursor's response to J5's second prompt was not that logically sound, and that the fix was not best practice. Seniors also criticized the juniors' prompt engineering habits, offering tips to prevent hallucination and inject useful contexts. S8 suggested that one of the juniors warm up their context and ask very specific tasks and questions to understand to the AI before giving bigger tasks. S10 stated that juniors should pay attention to the debugging steps suggested by tools (like adding console logs), even if they do not literally write them down and should use ask mode to inquire about ``typical causes'' rather than rely on agent mode for a single, specific solution.

Seniors also noted that they could make assumptions on the juniors based on their prompts. S6, seeing a junior prompt \emph{``What does handleDateRangeChange do? (adds Analytics.js as context)''} inferred: \emph{``I think they were feeling overwhelmed with the codebase.''} S10 observed that J6's natural-language prompting hinted at their familiarity with React, whereas in contrast, J8 relied on smaller technical prompts that often led them astray. S7 also mentioned that J4 seemed pretty comfortable with agent mode and that they mostly \emph{``did pay attention to what fixes were being made,''} accepting and reverting changes when necessary.

\subsubsection{Prompt history for mentorship and code review.}

\paragraph{\textbf{Seniors see prompt histories as efficient windows into thinking when pair programming is not feasible.}}
S10, S4, and S9 believe that prompt histories can give seniors a window into juniors' thinking without live pairing. Still, S6 and S1 prefer pairing (and even co-using the AI) when possible.

In terms of code review, S10 suggested that prompt review could be valuable for sharing examples of effective prompts that provide high-level information or narrow down problems faster, noting that \emph{``that kind of feedback in general just helps you use your tooling better.''} S8 noted that, due to poor tool history, their company tracks prompts in a shared Google Doc to surface useful inputs, and that a more structured prompt review document could be useful. S9 agreed, adding that prompts reveal developers' reasoning, while S7 disagreed, likening prompt scrutiny to questioning Stack Overflow use. 

S10 highlighted interest in understanding what developers do beyond prompting, citing J6 googling ``what are React state changes'' to provide \emph{``some context of where these prompts are coming from and trying to recreate their thought process.''} S7 added that they would want to know whether developers read Cursor's outputs and to see more granular timing data, explaining that \emph{``accept, accept, accept is a very different thing versus generating all that content and then not actually reading it''}

\section{Discussion}

\begin{quote}
 ``During the program life a programmer team possessing its theory remains in active control of the program, and in particular retains control over all modifications. The death of a program happens when the programmer team possessing its theory is dissolved.'' \cite{naur1985programming}
\\- Peter Naur, 1985
\end{quote}

\subsection{Before Agency is Dead: A Cautionary Pre-Mortem}

We discovered that agency in the software engineering workspace is configured \emph{before} the first prompt is sent: company policies, security regimes, and mandates to ``use AI weekly'' set default loci of control, and time pressure nudges engineers toward heavier automation \cite{denny2024prompt, Russo2024NavigatingGenAIAdoptionSE}. Within those constraints, along with the level of familiarity with the tools and the task, perceived capability-of self and of the tool-drives how much agency is ceded. When developers believe the tool is ``better than me,'' scrutiny drops; this helps on familiar terrain (safe acceleration with detailed design driven prompts or small basic tasks for both juniors and seniors) yet risks misdirection on unfamiliar terrain (scope creep, incorrect judgments, and brittle patches for juniors). This aligns with prior work on ``perception of AI's post‑human ability'' as a predictor of acceptance in the earlier stages and over-reliance when bad habits leads one astray \cite{gambino_2019_posthuman, lee2025impact}. Thus, we caution that building good habits of scrutinizing AI output comes first when interacting with AI tools \cite{buccinca2021trust, buccinca2025contrastive, kim2025fostering}. This responsibility falls on both junior and senior employees, organizations that impose AI-usage expectations and training \cite{zhang_2025_training}, as well as designers of human-centered AI tools for industry professionals \cite{yun2025generative, meem_2025_wellbeing}.

To remain autonomous is to be the author of one's reasons, not merely the approver of outputs. Juniors and seniors alike agreed that AI-generated code is still your own code, and that you must be accountable for it \cite{10.5555/3716662.3716672}. Juniors reported that authorship without understanding undermined fulfillment and ownership, while shaping the intent and using AI as labor---not judgment---restored achievement. Meeting engineers' metacognitive needs (hypothesis-setting, expectation-checking, evidence-seeking) is therefore essential to growth \cite{Pagliari2022SocialAgency, Bennett2023AgencyAutonomyHCI, EmirbayerMische1998, Darvishi2024Agency, bangerl2025creaitive}. Concretely, that means three non‑negotiables as a company wide and personal practice: (1) interruptibility/override, (2) legible provenance (actual sources, not just reasoning) alongside detailed verification, and (3) small, test‑bounded diffs---principles echoed in governance guidance for agentic systems and in studies on contextualized assistants \cite{Shavit2024AgenticPractices, Pinto2023ContextualizedAIAssistant}.

\subsection{Reframing the Talent Pipeline}

Industry narratives diverge: some firms cut junior roles as ``replaceable by AI,'' others prize ``AI-native'' hires \cite{npr2025jobs, wsj2025aisalaries}. Our study rejects both extremes. While juniors gained speed and exposure with AI tools, they reported fragile understanding and imposter feelings when comprehension lagged production, paralleling deskilling concerns \cite{weisz2025examining, Darvishi2024Agency, Choudhuri2024HowFarAreWe, Russo2024NavigatingGenAIAdoptionSE}. Workers risk becoming deskilled at core analytical tasks, summarizing information, finding sources, obtaining insights---while companies later hire humans to fix AI-generated details \cite{huseynova2024addressing, haque2025llms, kobiella2024if}. The traditional pipeline assumed gradual technical mastery through years of incremental complexity, but AI disrupts this progression. Our study reveals that junior and senior engineers adopt various strategies for using AI to support learning, with seniors drawing on prior, non-AI experiences to inform their use of AI tools, and juniors utilizing deliberate reflection and controlled use.

Thus, our findings suggest the pipeline must evolve, not disappear. The senior role transforms into ``Socratic guides and organizational anchors''---shifting from answering coding questions (which AI does well, and will likely continue to improve on) to asking the right questions that develop junior judgment. Seniors cultivate critical thinking through inquiry while providing irreplaceable context about organizational dynamics and unwritten rules that no AI can access. Junior growth reframes as ``earning judgment through deliberate restraint''---developing intuition for when not to delegate to AI, when to trust their instincts, and when to seek human guidance. The pipeline thus transforms from gradual responsibility increase to ``immediate accountability with guardrails''---juniors engage with production systems immediately but with accountability mechanisms like prompt reviews. We should view AI as another abstraction layer, similar to the transition from assembly to high-level languages \cite{10.1145/3715003}, where the next generation of seniors will be distinguished not by coding without AI, but by knowing when human insight matters most.

\subsection{A Suggested Practice: Prompt \& Code Reviews (PCRs)}

Our findings reveal a critical tension: junior engineers working with AI on unfamiliar tasks under time pressure experience loss of agency and ownership, reporting feelings of being ``fraudulent'' and losing their ``critical thinking muscle'' \cite{lee2025impact}. This erosion of agency stems partly from the increased opacity in AI-mediated work---when code emerges from prompts rather than direct authorship, accountability becomes diffuse. Based on our synthesis of 10 junior and 10 senior engineers' experiences, we suggest Prompt \& Code Reviews (PCRs) as a practice to restore agency through accountability. The generational divide makes this practice particularly relevant. Senior engineers developed foundational instincts in a pre-AI world---they understand what it means to own code from conception to completion. Junior engineers, entering as AI-natives, have never known professional software work without these tools. This asymmetry positions seniors uniquely to help juniors maintain agency by making AI interactions visible and accountable. PCRs operate under the principle that accountability preserves agency. When engineers must document and defend their prompting strategies, they remain authors of their reasoning even when AI generates the code. The practice would include: (1) problem framing and expected outcomes (human or AI-drafted), (2) key prompts and interaction modes that led to the solution, and (3) a brief human-written rationale for the chosen approach.

We recognize significant practical challenges in implementing PCRs. Code review is already time-intensive, and reviewing full prompt histories could overwhelm reviewers. Moreover, prompts often contain exploratory thinking and false starts that developers---particularly juniors---might feel uncomfortable exposing. Rather than advocating for exhaustive prompt audits, we envision PCRs as selective, lightweight artifacts focusing on critical decision points. Juniors could self-curate 2-3 key prompts that shaped their solution, while automation could filter out sensitive or exploratory content. PCRs represent a minimal intervention that addresses our core finding: agency in AI-mediated software engineering requires both individual accountability and organizational support. By making AI interactions part of the review process, teams can preserve the human judgment and ownership that both juniors and seniors identified as essential to meaningful work. In our study, seniors reviewing junior artifacts (even in brief 12-minute sessions) identified patterns of over-reliance, suggested better prompting strategies, and caught instances where juniors accepted incorrect AI suggestions. This visibility enables seniors to provide targeted mentorship---not just on code quality but on maintaining critical thinking when working with AI.

\subsection{Limitations \& Future Directions}

While our study provides exploratory insights into how engineers of different experience levels interact with AI in software development, several limitations warrant careful consideration. Our findings represent a snapshot in time (the summer of 2025) to help researchers understand software engineers at the cusp of widespread adoption of AI-assisted tools. Because our sample includes a diverse range of AI-adoption and experience levels, many of our insights are broadly generalizable to engineers at this time. Importantly, these limitations do not affect all findings equally, and considering their impact can help guide future research.

The differing structures of the study phases---where juniors and seniors completed different tasks---offer several opportunities for future research. While this design means the two groups cannot be directly compared in this paper, it provides exploratory insights that can motivate larger or deeper investigations \cite{bark2006use} into topics on agency in both generative and agentic AI, perceptions of career growth with AI, and the role of mentorship, compared across differing experience levels. Furthermore, although seniors and juniors did not engage in identical hands-on tasks, the task differences themselves reflect software engineering realities, where senior engineers tend to focus more on influencing project design and reviewing code, whereas junior engineers focus more on implementation and hands-on coding \cite{milewicz2019characterizing}. Additionally, both groups participated in semi-structured interviews with comparable probing questions to allow for high-level comparisons. This suggests that findings related to mentorship and AI as a mentor (e.g., both seniors and juniors suggesting AI as an accessible fallback first mentor) and findings on learning with AI (e.g., juniors navigating efficiency versus skill atrophy, seniors accelerating learning without replacing understanding) are likely robust across contexts, since both perspectives were captured in similar semi structured interview processes. However, behaviors tied to low and high familiarity (e.g., seniors' iterative refinement and process separation, juniors' constraint-based approaches and scope control issues) and AI tool usage patterns are more nuanced and less widely generalized, as seniors' reflections were largely retrospective, and juniors' behaviors were reflected upon with a short live coding task. Future work could build on this by developing parallel tasks or longitudinal comparisons.

The relatively small sample size and the concentration of participants in the USA and Canada point to opportunities to expand this research across more regions and organizational contexts. Despite this, the participants come from a diverse set of companies, which supports the ubiquity of findings around company-level norms (e.g, the pre-configuration of AI agency boundaries before individual preferences matter). Similarly, findings on career growth and seniority distinctions (e.g., seniors' foundational instincts guiding AI use, juniors' struggles with ownership and imposter syndrome) are likely widely applicable and reflect experiences reported across multiple organizations, aligning with broader industry discourse on AI adoption and skill development. While our strength lies in the depth of our qualitative, hour-long semi-structured interviews and diverse task based activities, future work could incorporate larger or more globally distributed samples to broaden representativeness.

Because the P1, P2, and P3 tasks were short-term and situational, future work could examine longer-term mentorship and professional development tasks. Such studies would allow for richer analysis of task behaviors and outcomes, rather than relying on these tasks as exploratory or needing to supplement them with additional data sources (e.g., interviews). Nonetheless, our bounded tasks complement the semi-structured interviews and survey responses, which emphasize longer-term developmental experiences related to mentorship and professional growth. In P1, senior engineers reconstructed their cumulative experiences, supported by ACTA to prompt more detailed reflection. In P2 and P3, the debugging and prompt-review activities functioned as proxy tasks to get a sense of their usual or organizational workflows~\cite{biehl2025prestige}. Across all phases, the tasks were designed to surface behaviors and insights and to prompt participants to reflect on their prior long-term experiences and practices without requiring a long-term study or long-term tasks. The study thus offers transferable insights into cognitive and interaction patterns that can inform future in-situ studies in organizational settings. In particular, the brief task of PCRs yielded valuable insights from senior engineers (e.g., start with understanding, use small testable batches, constant vigilance) as well as how they can infer junior thinking patterns, which point to the strong potential of PCRs in contemporary AI-mediated development environments.

Finally, while the controlled and time-constrained lab environment allowed consistency across participants, future work could explore how these activities unfold in naturalistic organizational settings. Although we used widely-used programming tools/frameworks (e.g., VSCode, React.js, Cursor), designed tasks informed by senior engineers, instructed participants to debug or review code as they normally would, and did not require them to think out loud or to use Cursor, in-situ studies would allow researchers to observe how these interaction patterns manifest in everyday development work across different tools and tasks that people might be more familiar with or might have more incentive to do well on.

In sum, future research could complement these findings with larger-scale and longer-term qualitative and quantitative studies on the operationalization and perception of agency in all areas (code generation, learning, and mentorship) in software to broaden generalizability. Future research should also evaluate optimal PCR formats, their impact on review burden, and their effectiveness in preserving the critical thinking skills necessary for professional growth in an AI-mediated environment.
\section{Conclusion}

Our study examines how agentic and generative AI integrate into software engineering, shaping agency, expertise, and mentorship across experience levels. While AI promises productivity gains, its impact on professional development is uneven: senior engineers leverage foundational instincts to maintain strategic control, whereas junior engineers gain exposure but risk fragile understanding and impostor syndrome. Crucially, AI cannot replace the tacit knowledge transfer essential for system thinking, failure anticipation, and architectural judgment. To address these challenges, we highlight three evolving practices. First, \textbf{Preserving Individual Agency}: company-wide and personal practices for using AI tools---such as incremental changes, interrupting, and verifying outputs---help engineers maintain control and responsibility over their work. Second, \textbf{Evolving the Mentorship Pipeline}: collaborative and individual practices in which senior engineers pass down intuition, critical thinking, and judgment to junior engineers support their learning, agency, and professional growth in an AI-mediated environment. Third, \textbf{Prompt \& Code Reviews (PCRs)}: a collaborative practice to maintain agency by making AI interactions accountable. Juniors document and justify key prompts, while seniors oversee accountability, ensuring AI-native juniors remain authors of their reasoning and retain ownership over outputs. Overall, preserving agency in software development, learning, and mentorship is essential to maintain a sustainable talent pipeline and ensure engineers can guide, not just operate, increasingly autonomous systems.

\begin{acks}
The authors would like to thank the anonymous reviewers, the senior and junior software engineers, and the many referrers who dedicated their time, insights, and or network for this study.
\end{acks}

\bibliographystyle{ACM-Reference-Format}
\bibliography{7-refs}

\onecolumn
\appendix
\section{Appendix}
\subsection{Phase 1 ACTA Brainstorming Results}
\label{sec:acta-brainstorm}

\begin{figure*}[h]
    \centering
    \includegraphics[width=0.75\textwidth]{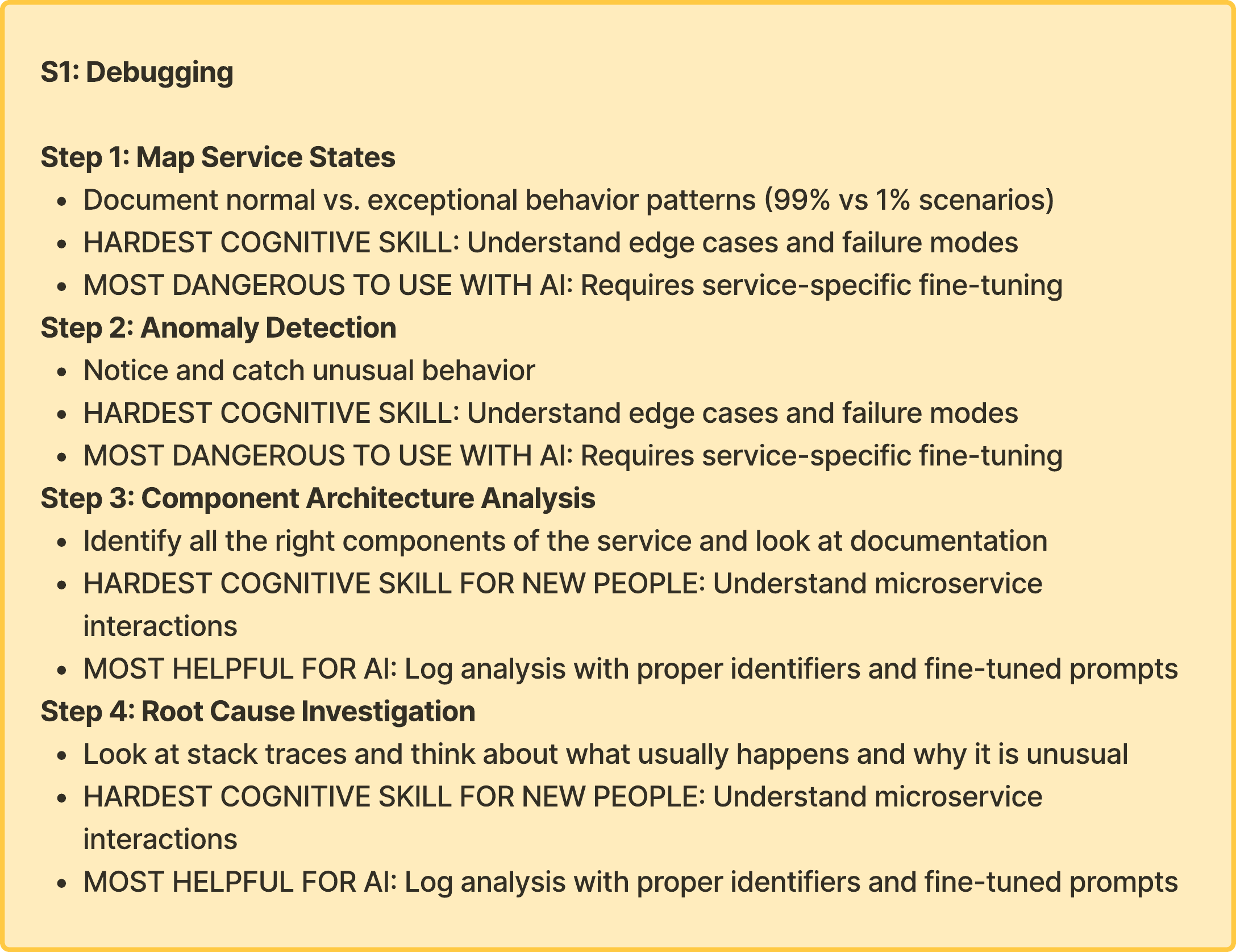} 
    \caption{\textbf{Senior engineer S1's Applied Cognitive Task Analysis (ACTA) decomposition of debugging workflows.} Task diagram showing S1's breakdown of debugging steps with annotations for cognitive demands and AI delegation considerations. S1's analysis illustrates how experienced engineers conceptualize the balance between maintaining control over critical decisions and leveraging AI for routine subtasks.}
    \Description{Task flow diagram for debugging workflow created by senior engineer S1. The diagram shows steps arranged in a workflow, with each step annotated with three types of labels: cognitive challenges, AI delegation opportunities, and AI delegation risks. The flow demonstrates how debugging progresses from initial problem identification through analysis to implementation and verification stages.}
    \label{fig:S1_ACTA}
\end{figure*}

\begin{figure*}[h]
    \centering
    \includegraphics[width=0.75\textwidth]{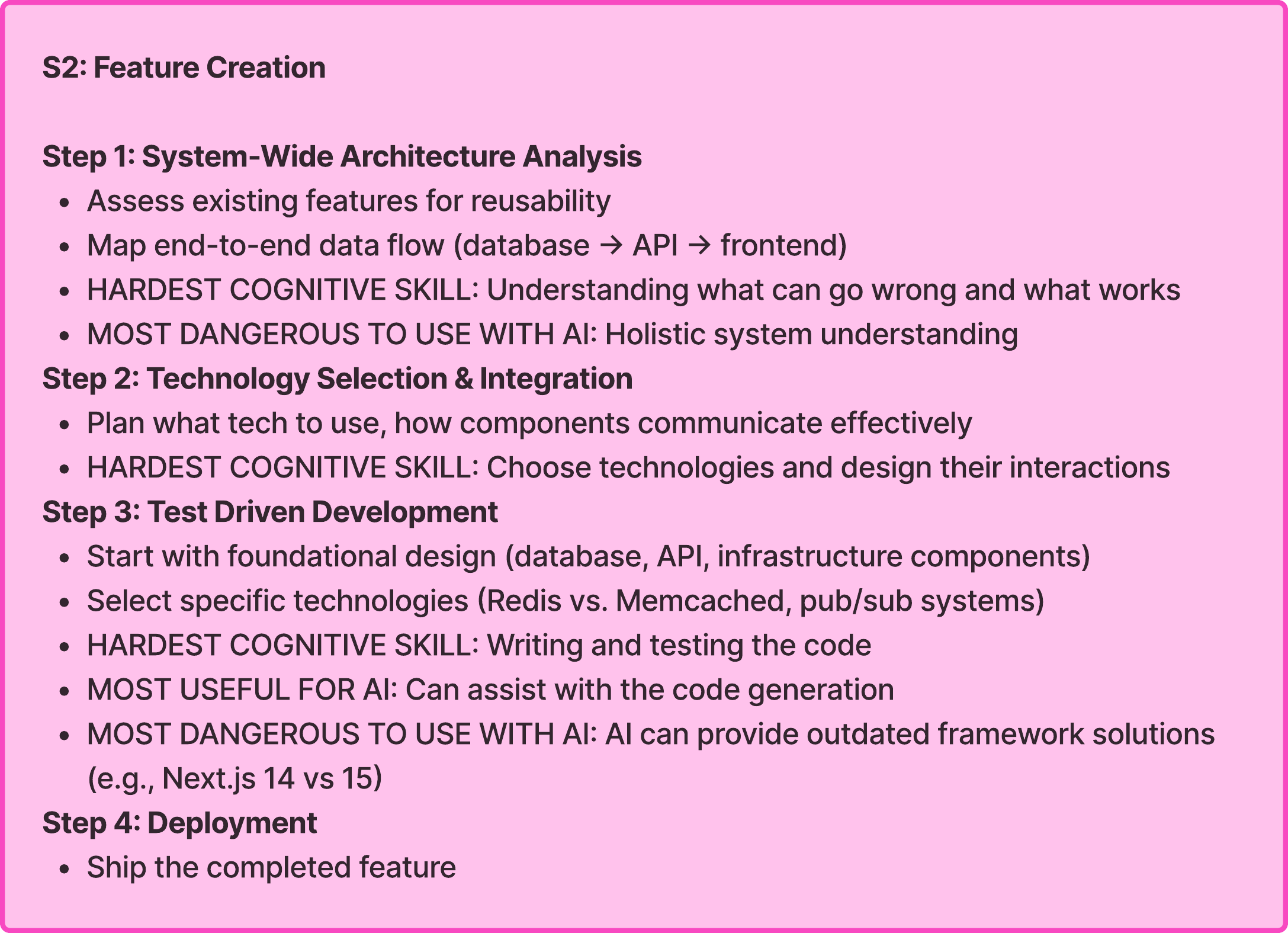}
    \caption{\textbf{Senior engineer S2's Applied Cognitive Task Analysis (ACTA) decomposition of feature creation workflows.} Task diagram illustrating S2's systematic breakdown of feature development processes with annotations for cognitive demands and AI delegation considerations. S2's perspective emphasizes the importance of human oversight in design decisions while acknowledging AI's utility in implementation tasks.}
    \Description{Task flow diagram for feature creation workflow created by senior engineer S2. The diagram shows steps arranged in a workflow, with each step annotated with three types of labels: cognitive challenges, AI delegation opportunities, and AI delegation risks. The flow progresses from initial requirements through design, implementation, testing, and deployment phases.}
    \label{fig:S2_ACTA}
\end{figure*}

\begin{figure*}[h]
    \centering
    \includegraphics[width=0.75\textwidth]{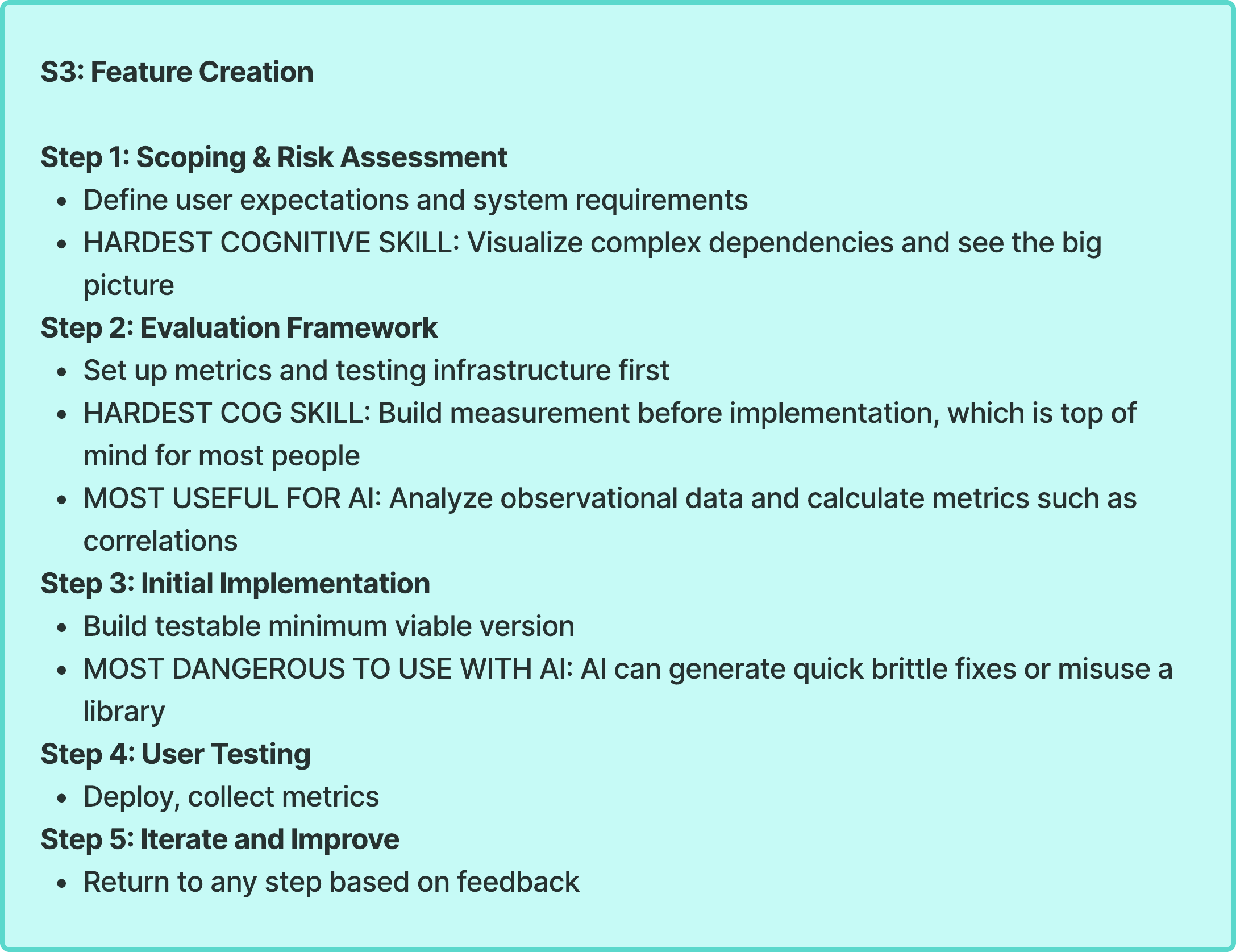} 
    \caption{\textbf{Senior engineer S3's Applied Cognitive Task Analysis (ACTA) decomposition of feature creation workflows.} Comprehensive task analysis showing S3's approach to feature development with explicit consideration of AI integration points with annotations for cognitive demands and AI delegation considerations. S3's analysis particularly emphasizes the role of system-level thinking and architectural coherence that AI currently cannot replicate.}
    \Description{Task flow diagram for feature creation workflow created by senior engineer S3. The diagram shows steps arranged in a workflow, with each step annotated with three types of labels: cognitive challenges, AI delegation opportunities, and AI delegation risks. The flow shows iterative cycles between design, implementation, and validation stages.}
    \label{fig:S3_ACTA}
\end{figure*}

\begin{figure*}[h]
    \centering
    \includegraphics[width=0.75\textwidth]{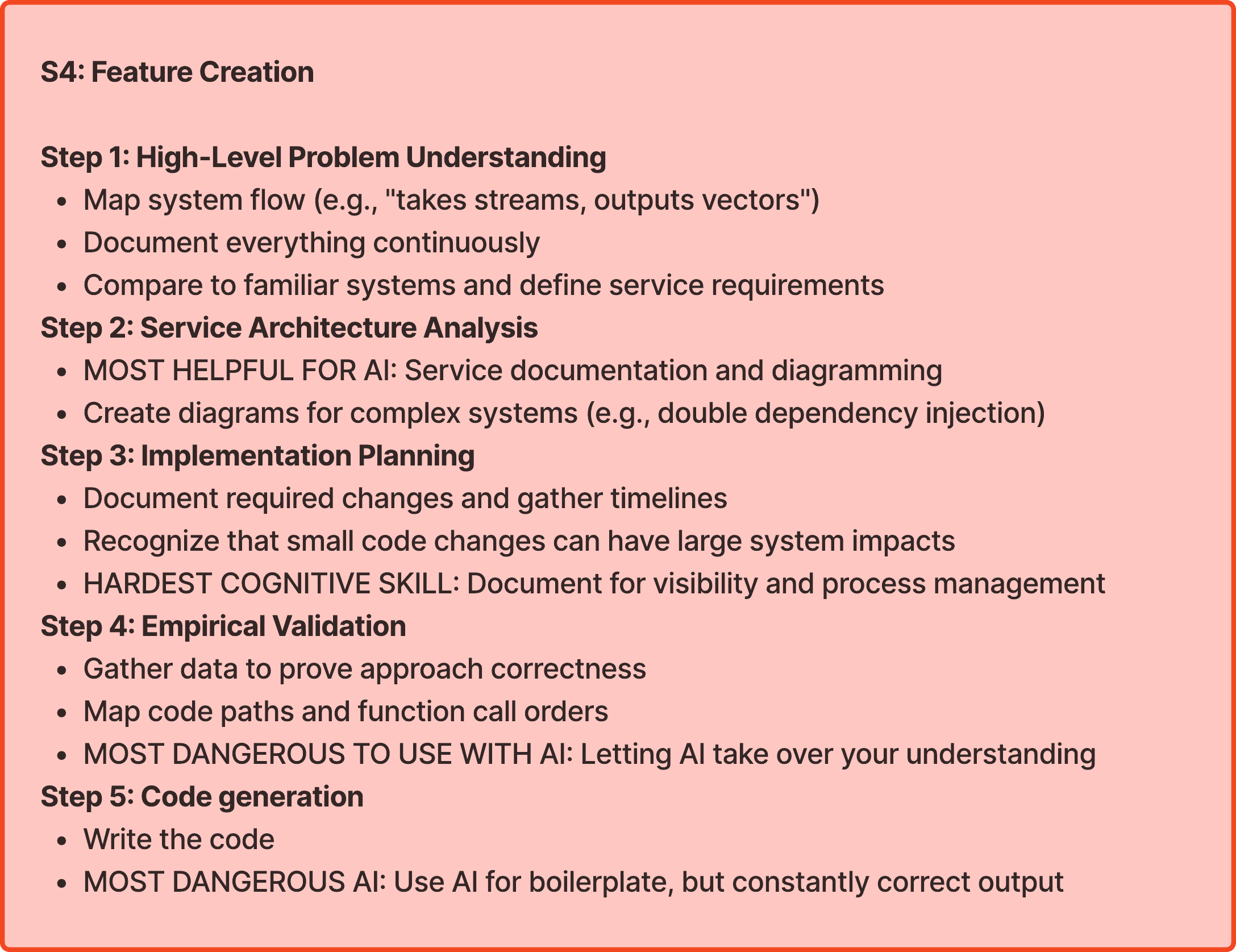} 
    \caption{\textbf{Senior engineer S4's Applied Cognitive Task Analysis (ACTA) decomposition of feature creation workflows.} Detailed cognitive task analysis revealing S4's nuanced understanding of feature development complexity and AI collaboration opportunities with annotations for cognitive demands and AI delegation considerations. S4's framework notably highlights the distinction between mechanical coding tasks suitable for AI and strategic decisions requiring human insight, reflecting years of experience in balancing automation with craftsmanship.}
    \Description{Task flow diagram for feature creation workflow created by senior engineer S4. The diagram shows steps arranged in a workflow, with each step annotated with three types of labels: cognitive challenges, AI delegation opportunities, and AI delegation risks. The flow emphasizes decision points and quality gates throughout the development process.}
    \label{fig:S4_ACTA}
\end{figure*}

\begin{figure*}[h]
    \centering
    \includegraphics[width=0.75\textwidth]{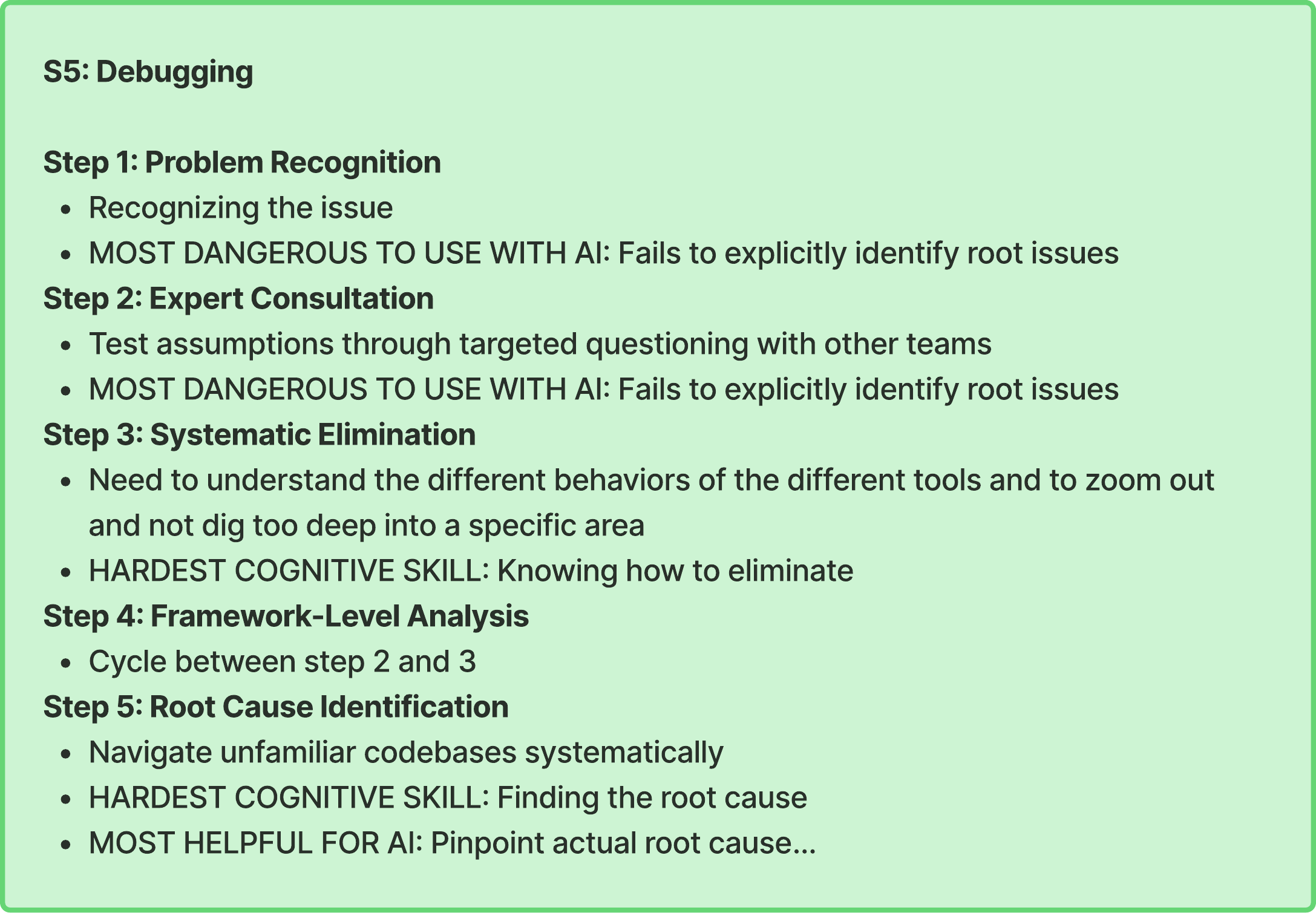} 
    \caption{\textbf{Senior engineer S5's Applied Cognitive Task Analysis (ACTA) decomposition of debugging workflows.} Systematic analysis of debugging processes through the lens of cognitive complexity and AI delegation potential with annotations for cognitive demands and AI delegation considerations. The analysis reflects S5's perspective on debugging as both a technical and investigative process, where AI serves best as an assistant rather than a replacement for human reasoning about system behavior and failure modes.}
    \Description{Task flow diagram for debugging workflow created by senior engineer S5. The diagram shows steps arranged in a workflow, with each step annotated with three types of labels: cognitive challenges, AI delegation opportunities, and AI delegation risks. The flow emphasizes the investigative nature of debugging with branches for different diagnostic paths and solution strategies.}
    \label{fig:S5_ACTA}
\end{figure*}


\end{document}